\documentclass[11pt,preprint]{aastex}
\usepackage{times}

\let\oldfootsep=\footnotesep
\newcommand\ltsima{$\; \buildrel <\over\sim \;$}
\newcommand\simlt{\lower.5ex\hbox{\ltsima}}
\newcommand\gtsima{$\; \buildrel >\over\sim \;$}
\newcommand\simgt{\lower.5ex\hbox{\gtsima}}

\newcommand\msun { \rm {M_\odot}}

\newcommand\lsun { \rm {L_\odot}}

\newcommand\umin{u_{\rm min}}

\newcommand\vperp{v_{\rm \perp}}
\newcommand\vp {\tilde{v}}
\newcommand\vpbold{\tilde{\bf v}}

\newcommand\pac{Paczy{\'n}ski }
\newcommand\ie{{\it i.e. }}
\newcommand\eclong{\lambda}
\newcommand\eclat{\beta}  
\newcommand\vs {{\bf v}_S}
\newcommand\vl {{\bf v}_L}

\newcommand\vhat{\widehat{v}}
\newcommand\that {\widehat{t}}

\newcommand\kms {\,{\rm km/s} }
\newcommand\kpc {\, {\rm kpc}}
 
\shorttitle{Microlensing Black Holes}
\shortauthors{Bennett et al.}

\begin{document}


\title{Gravitational Microlensing Events Due to
       Stellar Mass Black Holes\altaffilmark{1}}


\author{
    D.P.~Bennett\altaffilmark{2,3},
    A.C.~Becker\altaffilmark{4},
    J.L.~Quinn\altaffilmark{2},
    A.B.~Tomaney\altaffilmark{5},
      C.~Alcock\altaffilmark{3,6,7},
    R.A.~Allsman\altaffilmark{8},
    D.R.~Alves\altaffilmark{9},
    T.S.~Axelrod\altaffilmark{10},
    J.J.~Calitz\altaffilmark{11},
    K.H.~Cook\altaffilmark{3,7},
    A.J.~Drake\altaffilmark{7},
    P.C.~Fragile\altaffilmark{2},
    K.C.~Freeman\altaffilmark{10},
      M.~Geha\altaffilmark{12},
      K.~Griest\altaffilmark{13},
    B.R.~Johnson\altaffilmark{14},
    S.C.~Keller\altaffilmark{7},
      C.~Laws\altaffilmark{5},
    M.J.~Lehner\altaffilmark{6},
    S.L.~Marshall\altaffilmark{7},
    D.~Minniti\altaffilmark{15},
    C.A.~Nelson\altaffilmark{7,16},
    B.A.~Peterson\altaffilmark{10},
      P.~Popowski\altaffilmark{17},
    M.R.~Pratt\altaffilmark{5},
    P.J.~Quinn\altaffilmark{18},
    S.H.~Rhie\altaffilmark{2},
    C.W.~Stubbs\altaffilmark{3,5},
      W.~Sutherland\altaffilmark{19},
      T.~Vandehei\altaffilmark{12},
      D.~Welch\altaffilmark{20}\\
      {\bf (The MACHO and MPS Collaborations)}
        }

\clearpage


\altaffiltext{1}{Based in part on Observations from NASA's
                 Hubble Space Telescope}
\altaffiltext{2}{Department of Physics, University of Notre Dame, IN 46556}
\altaffiltext{3}{Center for Particle Astrophysics, University of California,
        Berkeley, CA 94720}
\altaffiltext{4}{Bell Laboratories, Lucent Technologies, 600 Mountain Avenue,
        Murray Hill, NJ 07974}
\altaffiltext{5}{Departments of Astronomy and Physics,
    University of Washington, Seattle, WA 98195}
\altaffiltext{6}{Department of Physics and Astronomy, University of
        Pennsylvania, Philadelphia, PA, 19104-6396}
\altaffiltext{7}{Lawrence Livermore National Laboratory, Livermore, CA 94550}
\altaffiltext{8}{Supercomputing Facility, Australian National University,
    Canberra, ACT 0200, Australia }
\altaffiltext{9}{Space Telescope Science Institute, 3700 San Martin Dr.,
    Baltimore, MD 21218}
\altaffiltext{10}{Research School of Astronomy and Astrophysics,
        Mount Stromlo Observatory, Cotter Road, Weston, ACT 2611, Australia}
\altaffiltext{11}{Physics Department,
    University of the Free State, Bloemfontein 9300, South Africa}
\altaffiltext{12}{Department of Astronomy and Astrophysics,
    University of California, Santa Cruz 95064 }
\altaffiltext{13}{Department of Physics, University of California,
    San Diego, CA 92039}
\altaffiltext{14}{Tate Laboratory of Physics, University of Minnesota,
  Minneapolis, MN 55455}
\altaffiltext{15}{Depto. de Astronomia, P. Universidad Catolica, Casilla 104,
        Santiago 22, Chile}
\altaffiltext{16}{Department of Physics, University of California, Berkeley,
        CA 94720}
\altaffiltext{17}{Max-Planck-Institut f\"{u}r Astrophysik,
        Karl-Schwarzschild-Str. 1, 85741 Garching bei M\"{u}nchen, Germany}
\altaffiltext{18}{European Southern Observatory, Karl Schwarzchild Str.\ 2,
        D-85748 Garching bei M\"{u}nchen, Germany}
\altaffiltext{19}{Department of Physics, University of Oxford,
    Oxford OX1 3RH, U.K.}
\altaffiltext{20}{Department of Physics and Astronomy, McMaster University,
    Hamilton, Ontario, Canada, L8S 4M1 }


\clearpage

\begin{abstract}
We present an analysis of the longest timescale microlensing 
events discovered by the MACHO Collaboration during a seven year survey of the
Galactic bulge. We find six events that exhibit very strong microlensing 
parallax signals due, in part, to accurate photometric data from the
GMAN and MPS collaborations. 
The microlensing parallax fit parameters are used
in a likelihood analysis, which is able to estimate the distance and
masses of the lens objects based upon a standard model of the Galactic
velocity distribution. This analysis indicates that the most likely
masses of five of the six lenses are $> 1\,\msun$, which suggests that
a substantial fraction of the Galactic lenses may be massive stellar remnants.
This could explain the observed excess of long timescale microlensing events.
The lenses for events MACHO-96-BLG-5 and MACHO-98-BLG-6 are the most massive,
with mass estimates of $M/\msun = 6{+10\atop -3}$ and 
$M/\msun = 6{+7\atop -3}$, respectively. The observed upper
limits on the absolute brightness of main sequence stars for
these lenses are $< 1\,\lsun$, so both lenses are black hole 
candidates. The black hole interpretation is also favored by a likelihood
analysis with a Bayesian prior using a conventional model for the lens
mass function.
We consider the possibility that the source stars for some of
these six events may lie in the foreground Galactic disk or in the Sagittarius 
(SGR) Dwarf Galaxy behind the bulge, but we find that bulge sources
are likely to dominate our microlensing parallax event sample.
Future HST observations of these events can either confirm 
the black hole lens hypothesis or detect the lens stars and 
provide a direct measurement of their masses. Future observations
of similar events by SIM or the Keck or VLTI interferometers \citep{vlti}
will allow direct measurements of the lens masses for stellar remnant 
lenses as well.
\end{abstract}




\section{Introduction}
\label{intro}

The abundance of old stellar remnants in our Galaxy is largely
unknown because they emit little radiation unless they happen to
be accreting material from a companion star, or for neutron stars,
if they happen to emit pulsar radiation in our direction. Gravitational
microlensing surveys \citep{liebes,pac86,macho-nat93,eros93,ogle93,moa01}
have the potential to detect completely dark
stellar remnants, but for most microlensing events, the mass can only
be estimated very crudely based upon the observed Einstein ring diameter
crossing time, $\that$. For an individual microlensing event, the mass
can only be estimated so crudely 
that a $7\msun$ black hole cannot be distinguished
from a $0.5\msun$ star. However, for some microlensing events, it is
possible to measure other parameters besides $\that$ that allow tighter
constraints on the lens mass \citep{refsdal-par,gould-par1,nem-fin,
macho-par1,macho-lmc9,hangould-xal,macho-95blg30,joint-98smc1,macho-96lmc2}.
For long timescale microlensing events,
which are often due to massive lenses, it is frequently possible to measure the
microlensing parallax effect \citep{refsdal-par,gould-par1,macho-par1} which is
an observable deviation in the microlensing light curve due to the orbital
motion of the Earth. In this paper, we present an analysis of the microlensing
events discovered by the MACHO Project which give a very strong microlensing
parallax signal, and we show that some of these events are best explained
as microlensing by black holes.

This paper is organized as follows: in Sections~\ref{sec-data} and
\ref{sec-longtime}, we discuss the microlensing event data set and the long 
timescale sub-sample. The microlensing parallax fits are presented in 
Section~\ref{sec-parfit}, and in Section~\ref{sec-mdest} we present
our main analysis to determine the distances and masses of the lenses.
This includes a discussion of the projected lens velocity distributions,
the source star color magnitude diagrams, and a likelihood analysis
of the distances and masses of the microlenses. In Section~\ref{sec-followup},
we discuss possible follow-up observations with high resolution telescopes
and interferometers that can directly determine the microlensing
parallax event lens masses, and we conclude in Section~\ref{sec-con}.

\section{The Data Set}
\label{sec-data}

The MACHO Project \citep{macho-nat93} has monitored $\sim 10-20$ million 
stars in the Galactic bulge for 6-7 months per
year during each of the 1993-1999 Galactic bulge seasons. During the
last half of 1994, real-time microlensing discovery with the MACHO Alert 
system became possible \citep{macho-alert}. (The OGLE collaboration
developed this capability the same year \citep{ogle-ews}.)
This development allowed much more accurate photometry of the 
microlensing events, which were discovered in progress, from the CTIO 0.9m 
telescope where the MACHO/GMAN Project was allocated about 1 hour every 
night \citep{becker-thesis}. In 1997, the Microlensing Planet Search (MPS) 
Project \citep{mps-98smc1} began microlensing follow-up observations from the
Mt.~Stromlo 1.9m telescope.
 
The data set used for this analysis consists of the MACHO survey data
from the Mt.~Stromlo 1.3m ``Great Melbourne" telescope for all seven
years, CTIO 0.9m data of selected alert events from 1995-1999, and
Mt.~Stromlo 1.9m data of alert events from the MPS 1997-1999 data sets.
The initial selection of events consists of 42 events from 1993
\citep{macho-bulge45}, 252 events discovered by the
MACHO Alert system from 1995-1999 (available from
{\tt http://darkstar.astro.washington.edu/}), and an additional 27 events
discovered during the testing of the alert system, for a grand total of
321 events. There are $\sim 200$ additional Galactic bulge
events that have been discovered via other analyses that we have not 
considered here. This paper will focus on the six events from
this list which give a strong microlensing parallax signal. The
coordinates of these events are given in Table~\ref{tbl-coords}.
A microlensing parallax study of a larger number of MACHO Alert
events is presented in \citet{becker-thesis}.

The MACHO and MPS data were reduced with slightly different versions of 
the SoDOPHOT photometry code \citep{macho-sod,macho-calib}. 
SoDOPHOT is quite similar
to the DOPHOT photometry code \citep{dophot} that it was derived
from, but SoDOPHOT photometry generally exhibits smaller photometric
scatter than DOPHOT photometry. This is due, in part, to SoDOPHOT's
error flags which allow the removal of suspect data points
\citep{macho-calib}, but the scatter in DOPHOT photometry is often
increased by the user's choice of PSF fitting parameters. Contrary to
expectations, allowing the PSF fit box size to scale with the seeing
generally causes increased photometric scatter \citep{macho-sod}.
The photometric errors reported by SoDOPHOT are modified by adding
of 1.4\% and 1.0\% in quadrature to the MACHO and MPS data,
respectively, to account for normalization and flat fielding errors.
The CTIO data were reduced with the ALLFRAME package \citep{allframe},
with the error estimates multiplied by a factor of 1.5 to account for
systematic errors. 

Table~\ref{tbl-nobs} gives the number of observations in each pass band
for each event, and the data
used for this paper are presented in Table~\ref{tbl-data}. The
complete set of macho survey data is available at 
{\tt http://wwwmacho.mcmaster.ca/} and {\tt http://wwwmacho.anu.edu.au}.
Only the MACHO survey data has been calibrated and transformed to
standard pass bands \citep{macho-calib}. The other data is given in
instrumental magnitudes which have only been calibrated relative to other
measurements with the same telescope and passband. 
The transformation between raw MACHO magnitudes given in Table~\ref{tbl-data}
($B_{\rm MACHO}$ and $R_{\rm MACHO}$)
and the Kron-Cousins $V$ and $R$ system is given by:
\begin{eqnarray}
   V & = & 2.699 + a\, B_{\rm MACHO} + b\, R_{\rm MACHO} \nonumber \\
   R & = & 2.412 + c\, B_{\rm MACHO} + d\, R_{\rm MACHO} \ ,
   \label{eq-calib}
\end{eqnarray}
where the coefficients, $a$, $b$, $c$, and $d$ are slightly different for
each event as shown in Table~\ref{tbl-calib}.

\section{Long Timescale Events}
\label{sec-longtime}

The timescale of a gravitational microlensing event is described by
the Einstein diameter crossing time, $\that$,  which depends on the lens
mass ($M$), distance ($D_\ell$), and transverse velocity ($\vperp$). It
is given by
\begin{equation}
   \that = {2 R_E\over \vperp} = {4\over \vperp c}
           \sqrt{GM D_\ell (D_s - D_\ell )\over D_s} \ ,
   \label{eq-that}
\end{equation}
where $D_s$ refers to the distance to the source (typically $8\,$kpc for a
bulge source), and $R_E$ is the radius of the Einstein Ring.
Eq.~(\ref{eq-that}) indicates that long $\that$ events can be caused by
large $M$, small $\vperp$ or both. Fig.~\ref{fig-thathist} shows the
long timescale tail of the $\that$ distribution for our sample of 321
Galactic bulge microlensing events. In their analysis of the
timescale distribution of the 1993 MACHO and OGLE bulge data sets,
\citet{hangould-mspec} noted a surprisingly large fraction
of the events with $\that \geq 140\,$days: 4/51 or 8\%. Such a large fraction
of long timescale events would be expected less than 2\% of the time
with any of the stellar mass functions that they considered. With our
data set of 321 events, we find 28, or 9\%, with $\that > 140\,$days.
The MACHO alert system is likely to be somewhat less sensitive to long
timescale events than the 1993 analysis because the alert trigger is based upon
the single most significant observation, so we would expect a slightly
smaller fraction of long timescale events, but the fraction reported here is
somewhat higher. The formal Poisson probability of 28/321 long events
when 2\% or less are expected is $< 10^{-10}$, so the excess of long
timescale events over the \citet{hangould-mspec} models is highly significant.
This disagreement may be due to a population of massive stellar remnants,
including black holes, that was not included in 
the \citet{hangould-mspec} models,
but there are other possibilities as well. Other explanations include
a set of source-lens systems that have a low relative velocity from
our vantage point in the Galactic disk or a more distant population
of source stars.

The microlensing parallax effect refers to the effect of the orbital
motion of the Earth on the observed microlensing light curve. The
photometric variation for most microlensing events lasts only
a month or two. For these events, the change in the Earth's
velocity vector during the event is too small to generate a detectable
deviation from the symmetric light curve, which is predicted for a constant
velocity between the lens and the Earth-source star line of sight.
For long timescale events, however, it is possible to see the effect
of the Earth's motion in the microlensing light curve, and this is
called the microlensing parallax effect. 

\section{Microlensing Parallax Fits}
\label{sec-parfit}

The magnification for a normal microlensing event with no detectable 
microlensing parallax is given by
\begin{equation}
  A(t) = {u^2 + 2 \over u\sqrt{u^2+4}} \ ;
\ \  u(t) \equiv \sqrt{u_0^2 + [2(t-t_0)/\that )]^2 } \ ,
   \label{eq-Astd}
\end{equation}
where $t_0$ is the time of closest approach between the angular positions
of the source and lens, and $u_0 = b/R_E$ where $b$ is the distance of the
closest approach of the lens to the observer-source line. Eq. (\ref{eq-Astd})
can be generalized to the microlensing parallax case
\citep{macho-par1} by assuming the perspective of an observer located at
the Sun. We can then replace the expression for $u(t)$ with
\begin{eqnarray}
       u^2(t) & = & u^2_0 + \omega^2 (t-t_0)^2
        + \alpha^2 \sin^2[\Omega(t-t_c)] \nonumber \\
& & + 2\alpha \sin[\Omega(t-t_c)]\,[\omega (t-t_0) \sin\theta
+ u_0\cos\theta ] \nonumber \\
& & + \alpha^2 \sin^2\eclat\,\cos^2[\Omega(t-t_c)]
+ 2\alpha \sin \eclat\,\cos[\Omega(t-t_c)] \,
        [\omega (t-t_0) \cos\theta - u_0\sin\theta]
   \label{eq-upar}
\end{eqnarray}
where $\eclong$ and $\eclat$ are the ecliptic longitude and latitude,
respectively,
$\theta$ is the angle between $\vperp$ and the North ecliptic axis,
$\omega = 2/\that$, and $t_c$ is the time when the Earth is closest
to the Sun-source line.
The parameters $\alpha$ and $\Omega$ are given by
\begin{equation}
\alpha = { \omega (1 {\rm AU}) \over \vp }
                \left( 1 - \epsilon \cos [\Omega_0 (t-t_p)]
\right) \ , \label{eq-al}
\end{equation}
and
\begin{equation}
\Omega (t-t_c) = \Omega_0 (t-t_c) + 2 \epsilon \sin[\Omega_0 (t-t_p)] \ ,
   \label{eq-Omega}
\end{equation}
where $t_p$ is the time of perihelion, 
$\Omega_0 = 2 \pi \,{\rm yr}^{-1}$, $\epsilon = 0.017$ is the
Earth's orbital eccentricity, and $\vp$ is
the lens star's transverse speed projected to the Solar position, given by
\begin{equation}
\vp = \vperp D_s/(D_s-D_\ell) \ , \label{eq-vp}
\end{equation}

\begin{figure}
\plotone{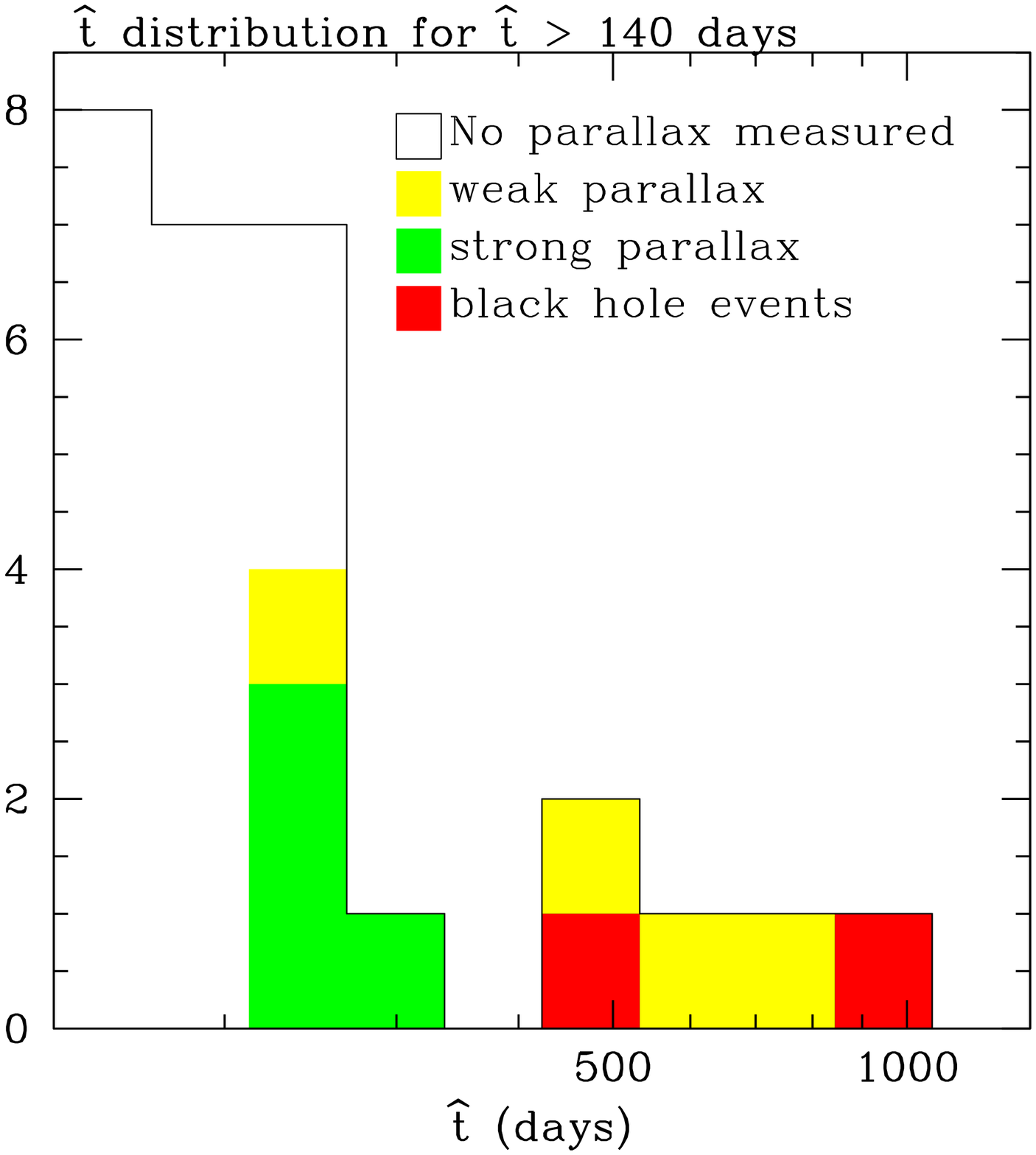}
\caption{
The distribution of event timescales, $\that$, for the
28 events with $\that > 140\,$ days. The colored bars indicate the events
with formally significant detections of microlensing parallax, but
parallax signal for the events indicated in yellow is weak enough that
the detection is not considered to be definitive.
\label{fig-thathist}}
\end{figure}

The 28 events shown in Fig.~\ref{fig-thathist} have been fit
with the microlensing parallax model described by 
eqs.~(\ref{eq-Astd})-(\ref{eq-al}) which has 5 independent parameters:
$t_0$, $u_0$, $\that$, $\vp$, and $\theta$. In the crowded fields that
are searched for microlensing, it is also necessary to include two
parameters for each independent photometric pass band (or telescope)
to describe the flux of the source star and the total flux of
any unlensed stars that are not resolved from the lensed source.
Thus, a microlensing parallax fit to the dual-color MACHO data alone will 
have 9 fit parameters, and a fit that includes the CTIO and MPS follow-up
data will have 13 fit parameters.

\begin{figure}
\plotone{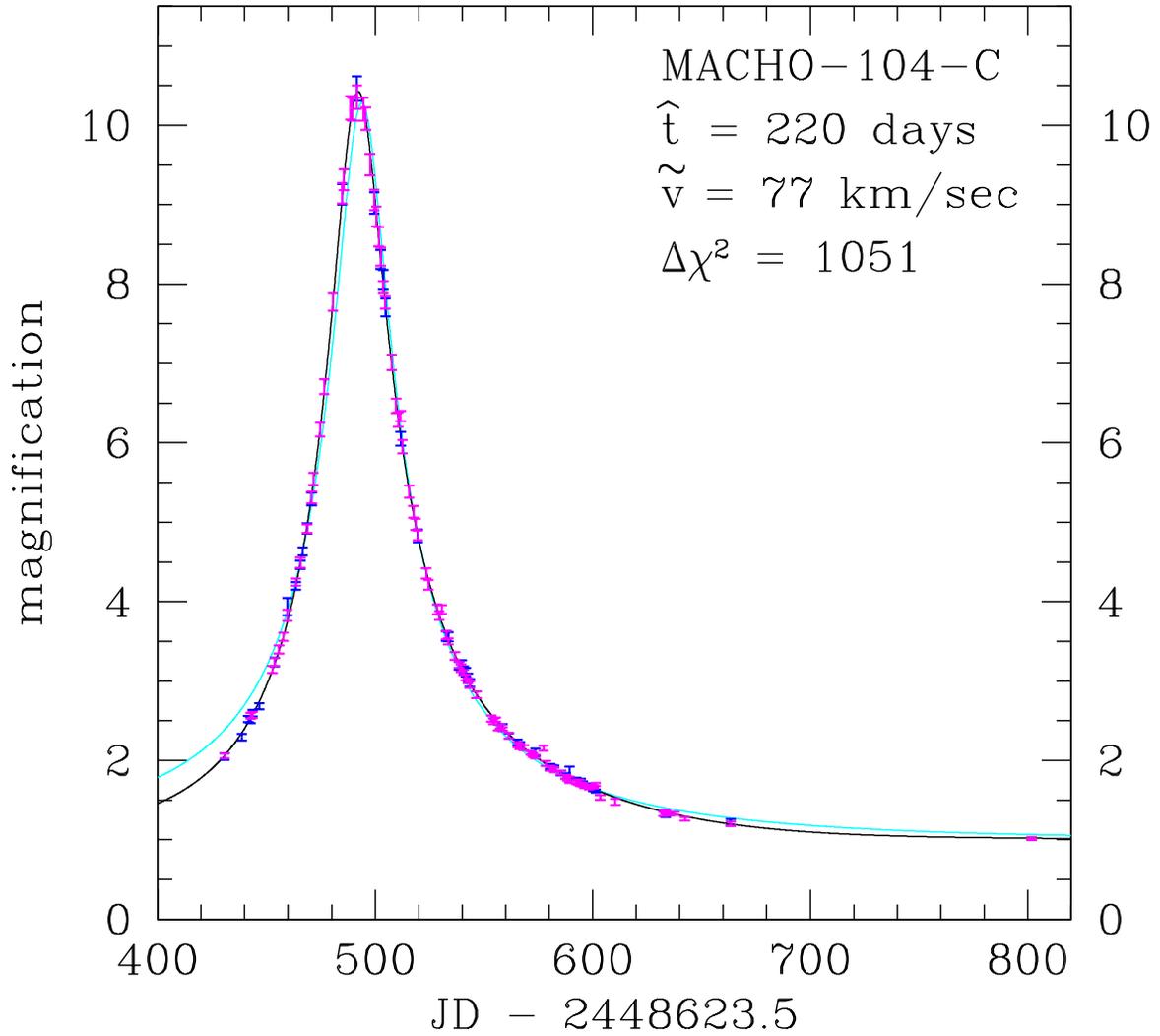}
\caption{
MACHO-104-C light curves normalized
to the unlensed flux of the lensed star. The MACHO red and blue data are
plotted in magenta and blue, respectively.
The black curve is the parallax fit while the cyan curve is the best fit
standard microlensing lightcurve. An additional 5 years of data showing
no photometric variation are not shown.
\label{fig-lc104c}}
\end{figure}

\begin{figure}
\plotone{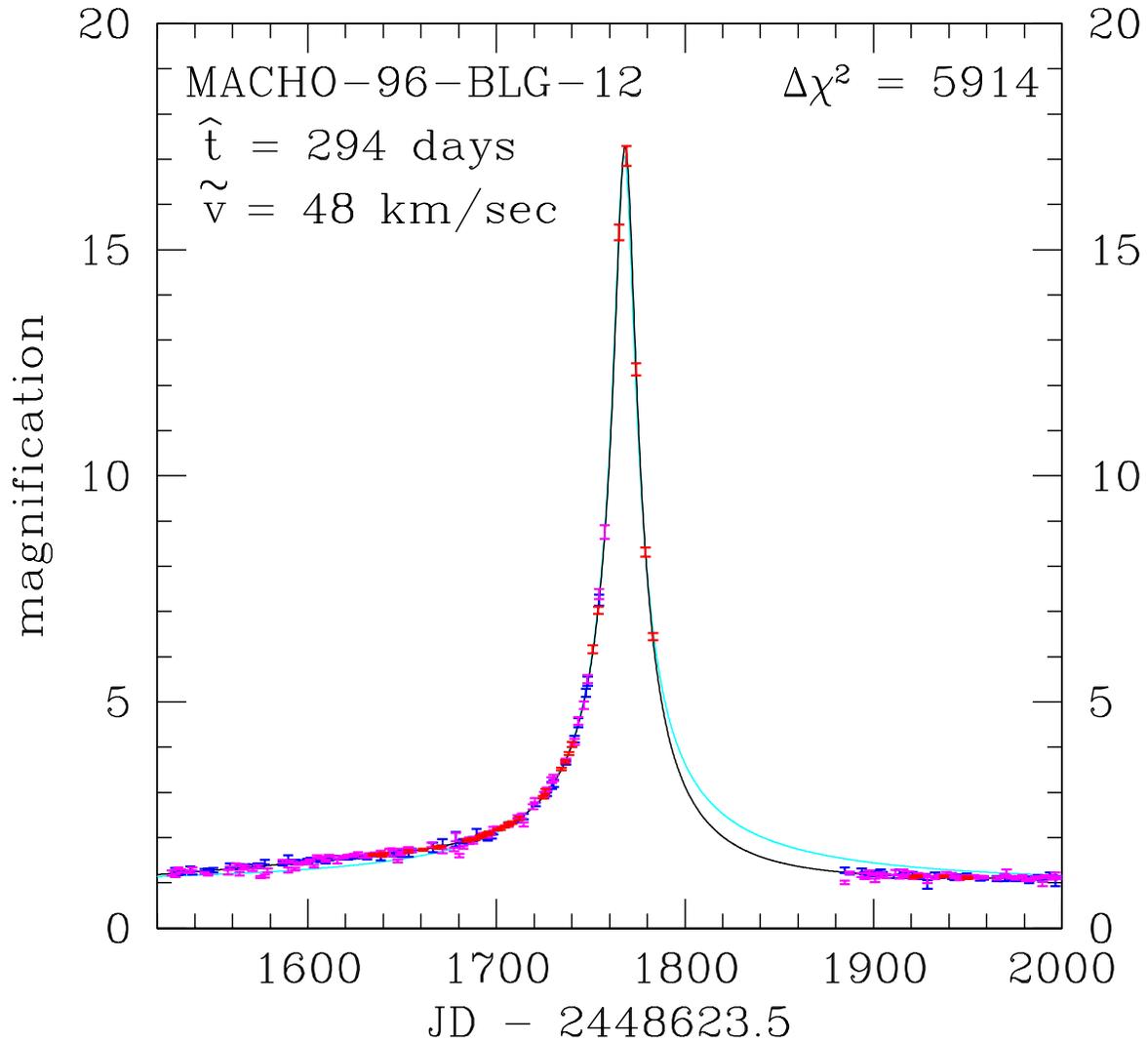}
\caption{
MACHO-96-BLG-12 lightcurve closeup with lightcurves normalized
to the unlensed flux of the lensed star. The MACHO red and blue data are
plotted in magenta and blue, respectively, and the CTIO data are shown in red.
The black curve is the parallax fit while the cyan curve is the best fit
standard microlensing lightcurve. An additional 5 years of data
showing no photometric variation are not shown.
\label{fig-lc96b12}}
\end{figure}

\begin{figure}
\plotone{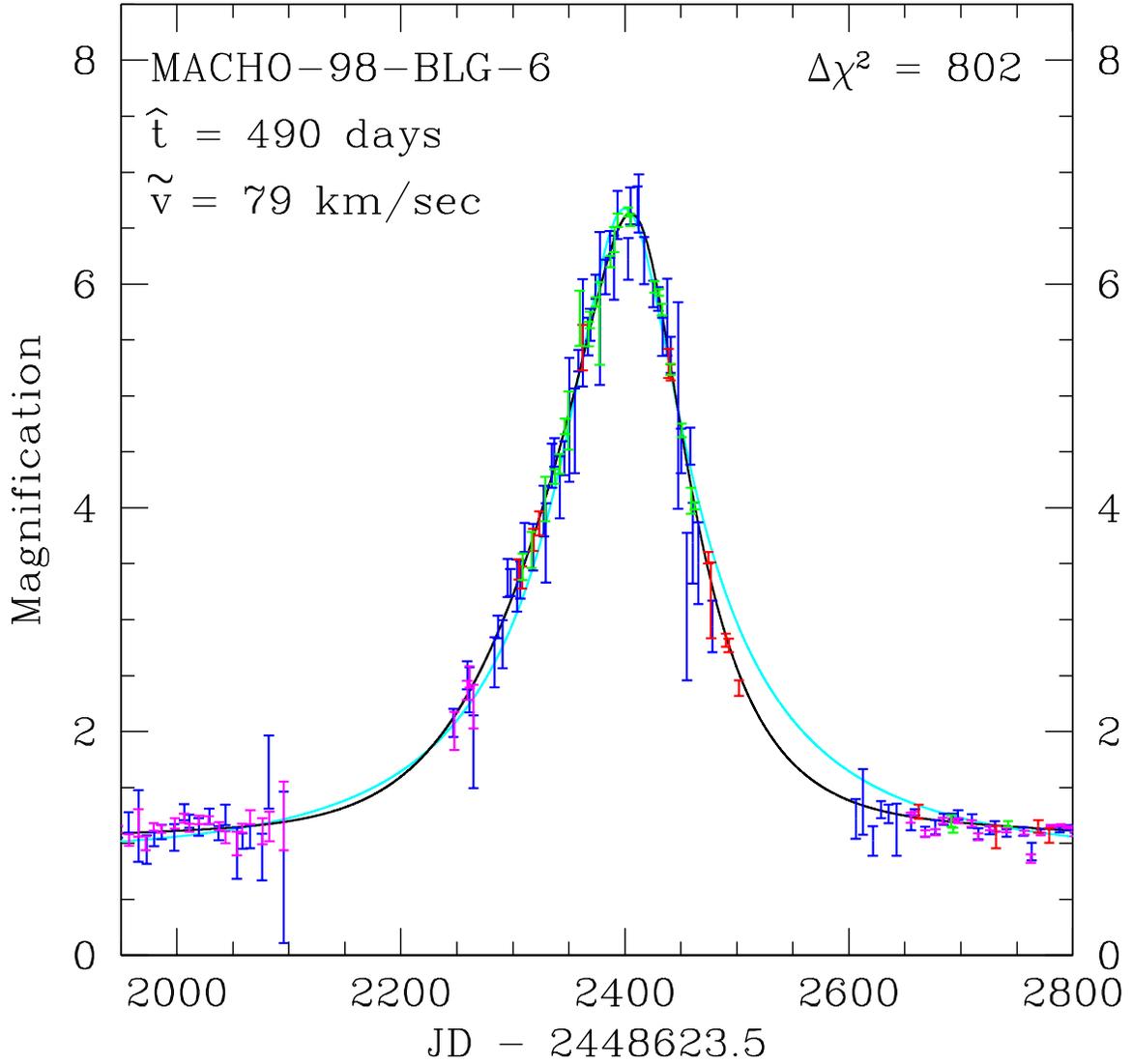}
\caption{
MACHO-98-BLG-6 lightcurve closeup with lightcurves normalized
to the unlensed flux of the lensed star. The MACHO red and blue data are
plotted in magenta and blue, respectively, the CTIO data are shown in red,
and the MPS data are shown in green.
The black curve is the parallax fit while the cyan curve is the best fit
standard microlensing lightcurve. The gap in the MACHO red data during the
day 2280-2650 interval is due to a CCD failure. An additional year of data
showing no photometric variation is not shown.
\label{fig-lc98b6}}
\end{figure}

\begin{figure}
\plotone{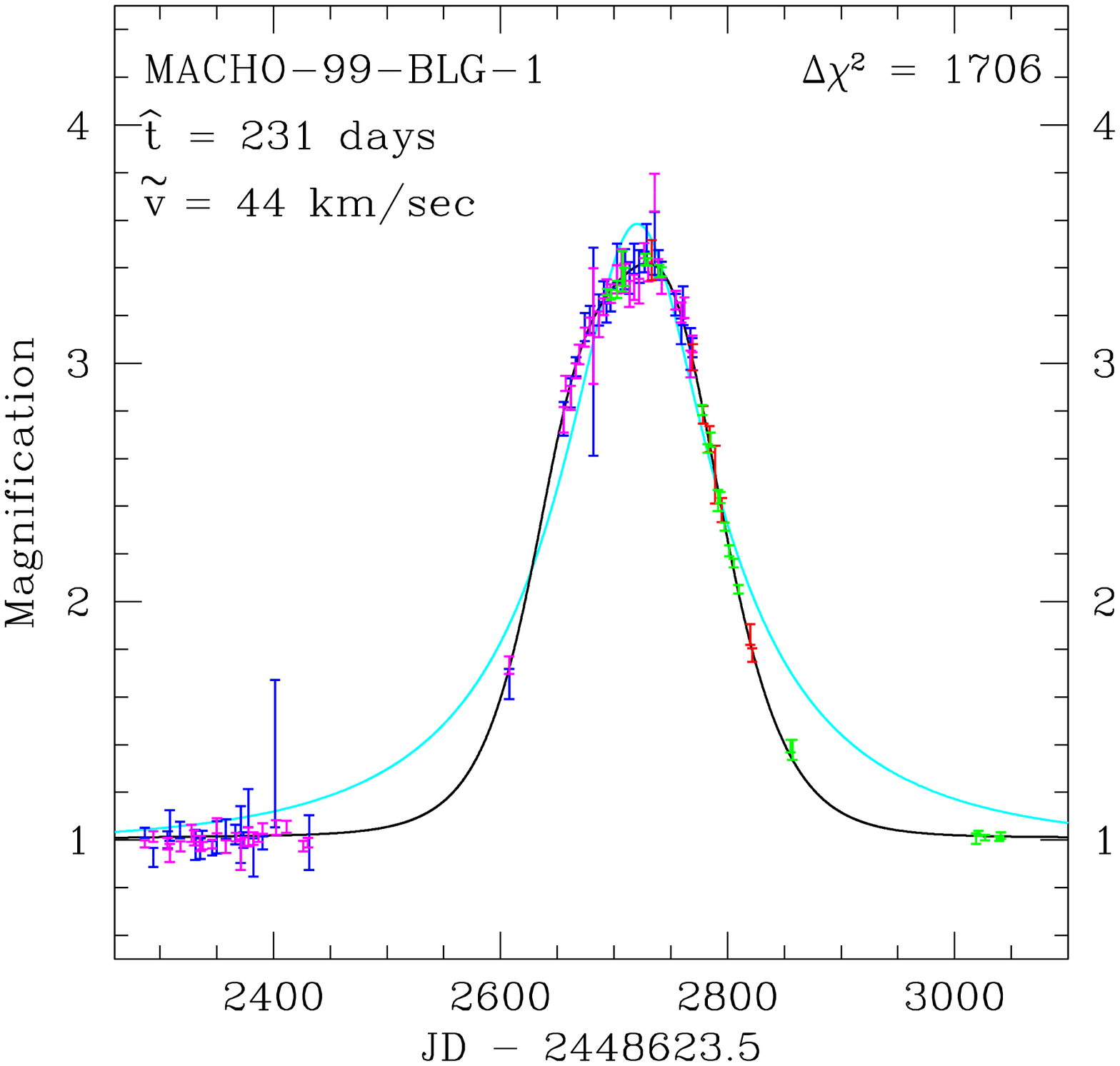}
\caption{
MACHO-99-BLG-1 light curves normalized
to the unlensed flux of the lensed star. The MACHO red and blue data are
plotted in magenta and blue, respectively, the CTIO data are shown in red,
and the MPS data are shown in green.
The black curve is the parallax fit while the cyan curve is the best fit
standard microlensing lightcurve. An additional 4 years of data showing
very little photometric variation are not shown.
\label{fig-lc99b1}}
\end{figure}

\begin{figure}
\plotone{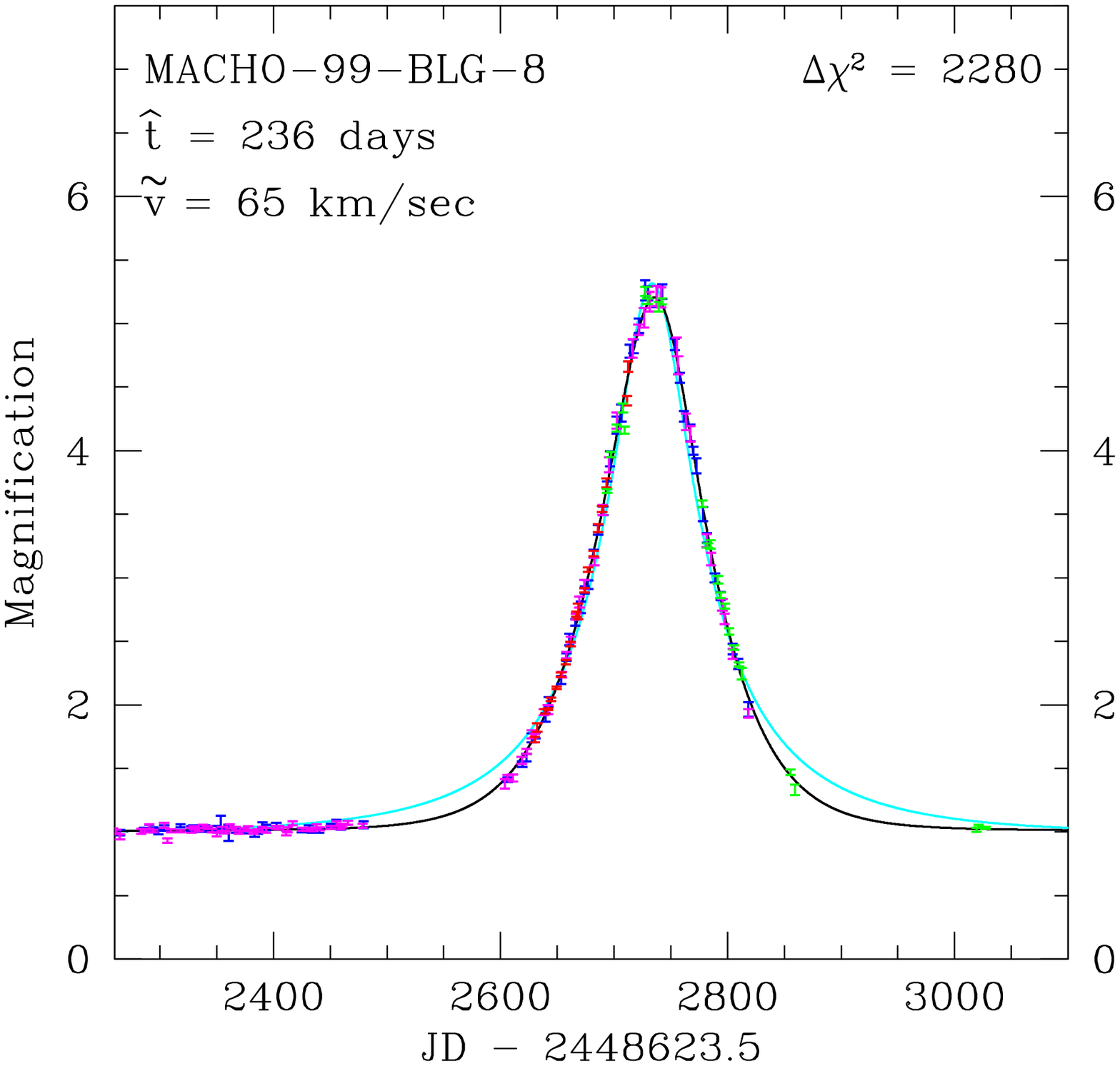}
\caption{
MACHO-99-BLG-8 light curves normalized
to the unlensed flux of the lensed star. The MACHO red and blue data are
plotted in magenta and blue, respectively, the CTIO data are shown in red,
and the MPS data are shown in green.
The black curve is the parallax fit while the cyan curve is the best fit
standard microlensing lightcurve. An additional 3 years of data showing
very little photometric variation are not shown.
\label{fig-lc99b8}}
\end{figure}

\begin{figure}
\plotone{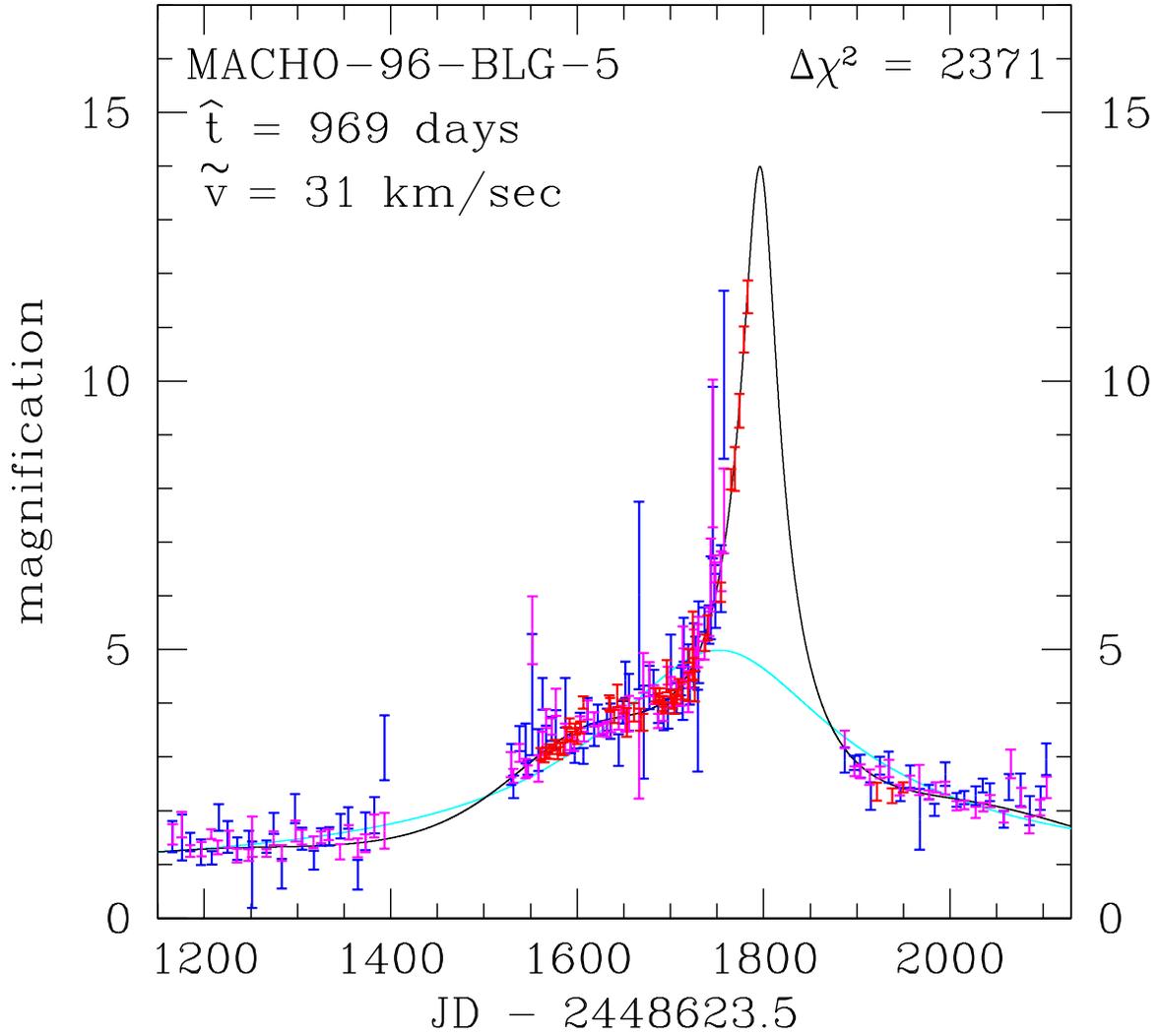}
\caption{
MACHO-96-BLG-5 lightcurves normalized
to the unlensed flux of the lensed star. The MACHO red and blue data are
plotted in magenta and blue, respectively, and the CTIO data are shown in red.
The black curve is the parallax fit while the cyan curve is the best fit
standard microlensing lightcurve. An additional 4 years of data showing
very little photometric variation are not shown.
\label{fig-lc96b5}}
\end{figure}

The microlensing parallax fits were performed with the MINUIT routine
from the CERN Library, and the results for the 6 events that we discuss in this
paper are summarized in Table~\ref{tbl-fitpar}. The best fit light curves
and data are shown in Figs.~\ref{fig-lc104c}-\ref{fig-lc96b5}. The 
significance of the microlensing parallax signal is represented by
the parameter $\Delta\chi^2$ shown in Table~\ref{tbl-fitpar} which is
the difference between the fit $\chi^2$ for a standard microlensing
fit with no parallax (\ie\ $\vp = \infty$) and the best fit presented
here. All 28 events with standard microlensing fits (including blending) which
indicated $\that_{\rm std} >140\,$days where fit with a microlensing parallax
model as well, and the 10 events with a microlensing parallax detection
with a significance of $\Delta\chi^2 \geq 50$ are indicated with color
in Fig.~\ref{fig-thathist}. The four events with 
$50 \leq \Delta\chi^2 < 200$ are 
MACHO-101-B, MACHO-95-BLG-27, MACHO-98-BLG-1, and MACHO-99-BLG-22, 
and the six strongest events with $\Delta\chi^2 \geq 200$ are 
MACHO-104-C, MACHO-96-BLG-5, MACHO-96-BLG-12, MACHO-98-BLG-6,
MACHO-99-BLG-1, and MACHO-99-BLG-8. 
These 6 events are the primary focus of this paper.

Note that most of the events
with $\that > 200\,$days and all the events with $\that > 300\,$days have
a significant parallax signal. Microlensing parallax is more easily 
detected in such long events because the Earth's velocity changes
significantly during the event, and because long events are likely to
have low $\vperp$ values. There are also a number of events with
much shorter timescales that appear to have microlensing parallax
signals significant at the $\Delta\chi^2 \geq 50$ level, but many of 
these have rather implausible parameters. This is likely to be due to the
fact that other effects besides microlensing parallax can perturb the
microlensing light curves in ways that can mimic the parallax effect.
Examples of this include binary microlensing (see the discussion of
MACHO-98-BLG-14 in Alcock et al. 2000a),
and the reverse of the parallax effect, the orbital motion of a binary
source star \citep{eros-sparm2,macho-96lmc2,griest-hu,hangould-xal}, 
sometimes called the
``xallarap" effect. The xallarap effect can be particularly difficult to
distinguish from microlensing parallax because a xallarap light curve can be
identical to a parallax light curve if the period, inclination, eccentricity,
and phase mimic that of the Earth. In practical terms, this is a
difficulty only when the xallarap or parallax signal-to-noise is weak 
so that the fit parameters are poorly determined.

In order to avoid contamination of our microlensing parallax sample
with non-parallax microlensing events, we have set a higher threshold
for the events that we study in detail in this paper: $\Delta\chi^2 \geq 200$.
The 6 events that pass this threshold are listed in Tables~\ref{tbl-coords}
and \ref{tbl-fitpar}. One of these events, MACHO-104-C, was the
first microlensing parallax event ever discovered \citep{macho-par1},
and the other five events were discovered by the MACHO Alert system.
Because of this, they had the benefit of follow-up observations by
the MACHO/GMAN Collaboration on the CTIO 0.9m telescope or 
by the MPS Collaboration on the Mt.~Stromlo
1.9m telescope. Four of these five events would have passed the
$\Delta\chi^2 \geq 200$ cut without the follow-up data, but event
MACHO-98-BLG-6 only passes the cut because of the MPS follow-up data.
This is probably due to a CCD failure that prevented the imaging of this
event in the MACHO-Red band during most of the 1998 bulge season.

We have also compared our microlensing parallax fits to binary lens fits
for each of these events. The parallax fits are preferred in every case
with $\chi^2$ improvements of 70.3, 81.6, 1625.8, 227.2, 1957.2, and
1601.3 for events MACHO-104-C, MACHO-96-BLG-5, MACHO-96-BLG-12,
MACHO-98-BLG-6, MACHO-99-BLG-1, and MACHO-99-BLG-8, respectively.

\subsection{HST Observations of MACHO-96-BLG-5}
\label{sec-96b5}

Event MACHO-96-BLG-5 is both the longest event in our 
sample\rlap,\footnote[1]{The analysis of the MACHO and MPS data for
MACHO-99-BLG-22 gives a best fit $\that = 700\,$days, but a 
combined analysis with the OGLE data yields 
a fit that is similar to the OGLE result \citep{ogle-99b22} and
gives $\that = 1100\,$days. This is about 10\% 
longer than our result for MACHO-96-BLG-5.} 
and the event with the faintest source star. In fact, the 
microlensing parallax fit does not
constrain the source star brightness very well. This is due to the
faintness of the source, and due to a potential systematic error.
The MACHO camera had a CCD upgrade in early 1999 which put a new
CCD in the location that views MACHO-96-BLG-5 in the MACHO-Red passband.
The new CCD probably had the effect of shifting the effective 
bandpass to a slightly different central wavelength, and so a slight systematic
shift in the photometry of all the stars might be expected to occur
with this upgrade. Because the MACHO-96-BLG-5 source is strongly blended
with unlensed neighbors, the effect of this slight shift on the 
microlensing fit parameters can be relatively large because the 
fitting routine tries to explain all flux variation as resulting from
microlensing. The best fit, with this suspect data removed, indicates that
only about $12\pm 3$\% of the flux associated with the ``star" seen in our
ground-based images has been microlensed, which would imply that the
remaining 88\% of the flux must come from unlensed neighboring stars
which are within $\sim 1.5$" of the lensed source.
Fortunately, we have a set of images from the Hubble Space
Telescope's WFPC2 Camera that can be used to constrain the brightness of
the source star more accurately than the fit does.

We had one orbit of HST data taken in the V and I (F555W \& F814W)
passbands of the WFPC2 Camera through Director's Discretionary Proposal
\# 8490, and this can be used to identify the microlensed source star.
The first step in this identification process is to determine the centroid
of the star that was lensed. This can be accomplished by subtracting
two images which have substantially different microlensing magnifications
\citep{macho-hstlmc1}.
Since it is only the lensed source star that will appear to vary in
brightness, this procedure will yield a point source centered on the location
of the lens and source. Of course, the subtraction procedure must take
into account the differences in the observing conditions of the two
frames, including differences in seeing, pointing, sky brightness and air mass.
We have accomplished this with the use of the DIFIMPHOT package of
\citet{tomaney-dip}. 

\begin{figure}
\plottwo{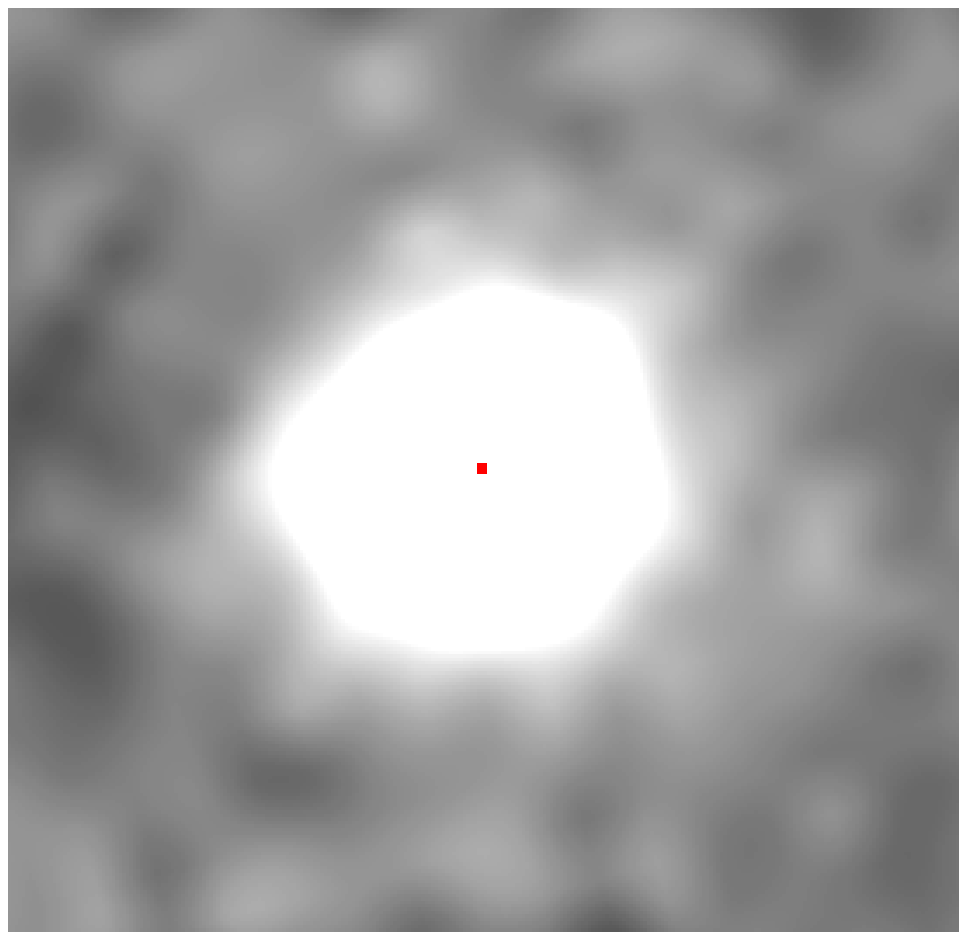}{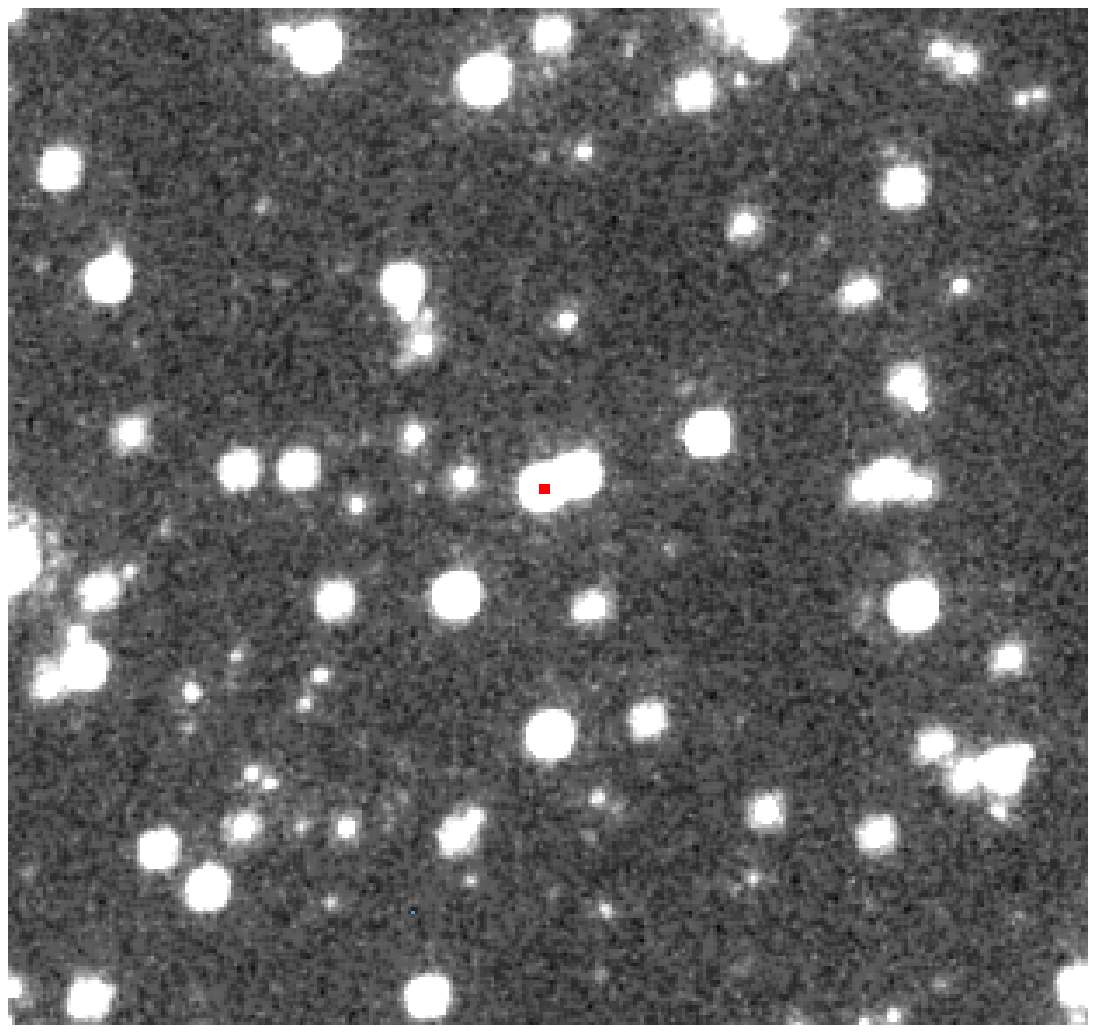}
\caption{
The image on the left is the master difference image as described in the
text. It has been registered to the same coordinate system as the
F814W HST/WFPC2 image shown on the right. The red marks show the 
centroid of the variable flux in the master difference image and the
location of this centroid when transformed to the coordinate system of
the HST data. A single, main sequence bulge star is clearly identified
as the lensed source star.
\label{fig-hst}}
\end{figure}

A set of 18 of our best CTIO images were selected to
use for this source location task because the CTIO images generally have
better seeing than the MACHO images, and because the highest magnification of
the source was only observed from CTIO. These 18 images were combined to
construct a master reference image which was then subtracted from each
individual frame to construct a set of 18 difference images. The difference
frames which had a negative flux at the location of our target were
inverted, and then all the difference images were combined to make
the master difference image shown in Figure~\ref{fig-hst}. The centroid
of the excess flux in this master difference image can be determined to
better than 0.01\rlap."

In order to identify the lensed source star on the HST images, we must
find the correct coordinate transformation to match the ground and HST
frames, but this is complicated by the fact that most of the ``stars"
in the ground-based images actually consist of flux from several different
stars that are blended together in the ground-based frames. We have dealt
with this in two different ways: first, we 
used the HST images to select a list of stars
that were much brighter than their near neighbors, so that their positions
should not be greatly affected by blending in the ground-based images.
Then, we convolved the HST data with a 1.2" FWHM Gaussian PSF to simulate the
resolution of the ground-based CTIO data. We then analyzed the convolved
HST image with the same data reduction software used for the ground-based
data. This gave an additional star list from the HST image. Two
independent coordinate transformations between the ground-based 
and HST data were obtained by matching these two stars list to the
star list for the ground-based data.

The HST images were dithered, and we combined them
with the Drizzle routine \citep{drizzle} prior to the comparison
with the ground-based data.  The CTIO R-band data were compared
to the HST I and V band images as well as sum of the I and V band images.
Coordinate transformation were determined to match the CTIO image
coordinates to each of these HST images using the bright, isolated
stars in the HST images and with the HST images convolved to ground-based
seeing. This resulted in a total of 6 different comparisons
between the location of the lensed star in the CTIO image and the HST images. 
All these comparisons
yielded the same lens star location on the HST frames to better than
0.02", and this location coincides with the centroid of the star indicated
in Fig.~\ref{fig-hst}. This star was examined carefully in both the V and I
images to determine if it could be a blend of more than one star. Model
and DAOPHOT generated PSFs were subtracted at the centroid
location of the lensed source, but no hint of any additional star 
was found. This star is very likely to be the
source star for the MACHO-96-BLG-5 microlensing event.

The next step in the comparison of the HST and ground-based data is to
determine what fraction of the flux of the object identified as a star
in the ground-based frames is contributed by the source star identified
in the HST images. This task is complicated by the fact that there is
no close correspondence between the passbands of the ground based images
and those used for the HST data. (This is due to the limitations imposed
upon an HST Director's Discretionary time proposal. We requested prompt
images in V and I to confirm the photometric variation implied by the
microlensing parallax model, but prompt imaging in R could not be justified.
Imaging in R was obtained in a subsequent GO program, and the analysis
of these data will appear in a future publication.)
Presumably, some combination of the V and I band images would provide
a good representation of the R band ground based image.

The determination of the lensed flux fraction was made as follows:
Photometry of the V+I combined HST frame was obtained using the IRAF
implementation of DAOPHOT \citep{daophot} and also using SoDOPHOT.
Both of these packages were also used to reduce the HST images which
had been convolved to mimic ground based seeing. The
total stellar flux of isolated, bright stars was not conserved in
these convolved images, so we found it necessary to renormalize the
stellar flux in the convolved images to the ratio found for these 
isolated bright stars. This comparison yielded a flux fraction of
36\% for the lensed component of the stellar blend identified as a single
star in the ground-based images.  We also followed this same procedure for the
separate I and V images, and the results for the lensed flux fraction
were quite similar as might be expected from the fit results shown in
Table~\ref{tbl-fitpar}, which indicate no color dependence for
the blending fit parameter. This is likely to be due to the fact that the
stars contributing the blended light and the lensed source star are all
main sequence stars of similar color which are
just below the bulge turn-off.

We must also make a correction for the fact that the source star was
still being magnified by the lens when the HST frames were taken. Because
the event timescale depends upon the amount of blending that we determine
from the HST analysis, it requires an iteration or two to find a fit
that predicts the observed brightness of the lensed star in the HST frames.
The best fit result is that the lensed source provides 33\% of the
total flux of the blended object that would be seen in the ground based
frames in the absence of any microlensing magnification. At the time of
the HST images, the lensing magnification was 1.063 according
to this fit.

Finally, we should mention the possibility that the star identified
with the lensed source centroid is not, in fact, a single star. The HST
images reveal no evidence of a chance superposition of unrelated stars, so
this is unlikely. However, it could be that the superposition is not
due to chance. Suppose, for example, that the star we've identified as the
MACHO-96-BLG-5 source is actually the superimposed images of the lens and 
source.  While this is a logical possibility, we will show below that there is
no plausible scenario for this to occur because the implied lens mass
cannot be made compatible with the observed brightness of the lens plus
source.

\section{Lens Mass and Distance Estimates}
\label{sec-mdest}

The measurement of the projected speed of the lens, 
$\vp$, allows us to relate the
lens mass to the lens and source distances
\begin{equation}
   M = {\vp^2 \that^2 c^2 \over 16 G} {D_s-D_\ell \over D_\ell D_s}
     = {\vp^2 \that^2 c^2 \over 16 G} {1-x \over x D_s} \ .
   \label{eq-m}
\end{equation}
It is often assumed that the distance to the source, $D_s$, is already
known, at least approximately, so this relation can be considered
to give the lens mass as a function of distance. Given the lens distance,
one can also work out the lens velocity with respect to the line-of-sight
to the source, $\vperp$. But, for some distances, the implied $\vperp$
value can be unreasonably small or large. Thus, with some knowledge of
the Galactic velocity distribution, we can work out an estimate
for the distance and mass of the lens. This has been done for the
MACHO-104-C event using a likelihood method in \citet{macho-par1}. This
analysis assumes that the source star resides in the Galactic bulge,
which is true for the vast majority of microlensing events seen towards the
Galactic bulge.  The result of similar analyses for the events
presented in this paper are summarized in Table~\ref{tbl-mass}
and in Figs.~\ref{fig-masslike-bh}-\ref{fig-masslike-99}.
However, the microlensing parallax events are selected
from a sample of unusually long microlensing events, so it may be
that their source star locations are atypical as well.
With the data currently available to us, we have two ways to
investigate the location of the source stars for our microlensing
parallax events. The first is to make use of the direction of 
projected velocity as determined by the microlensing parallax fit, and
the second is to examine the location of the source star in a
color-magnitude diagram of nearby stars. 
Another, perhaps more effective, discriminant between different source 
populations is radial velocity measurements.  Radial velocities for 
some of the source stars have been measured by \citet{cook-spectra}, 
and they have provided us with some preliminary results.

\subsection{Source Star Locations}
\label{sec-sourceloc}

The line-of-sight toward a Galactic bulge microlensing event passes through
the Galactic disk, the bulge, and through the Sagittarius (SGR) Dwarf Galaxy
behind the bulge. So, all of these are possible locations for the source
stars. The variation in the source population/location can affect the
inferred properties of the lens in several different ways:
\begin{enumerate}
   \item Microlensing rate: The microlensing rate per source star
         is very much lower for
         foreground Galactic disk stars and very much higher for SGR Dwarf
         stars than for Galactic bulge stars. Thus, foreground disk stars
         and SGR Dwarf stars will be under-represented and over-represented,
         respectively, in samples of microlensed stars when compared to
         stars in the Galactic bulge.
   \item Microlensing parallax detectability: some source star populations
         such as the foreground Galactic disk and the Sagittarius Dwarf
         Galaxy give rise to a larger fraction of events with microlensing
         parallax parameters that can be measured.
   \item Source distance: A source at a greater distance than the nominal
         Galactic bulge distance will usually imply a lower lens mass
         since $M$ is a decreasing function of $D_s$ in eq.~(\ref{eq-m})
         (for fixed $x$).
         Similarly, a smaller $D_s$ implies a larger mass.
   \item Source velocity: From eq.~(\ref{eq-vp}), we see that, for a fixed $\vp$
         value, a smaller $\vperp$ value implies a smaller $D_s$ which, in 
         turn, implies a smaller lens mass (for fixed $D_\ell$).
         Smaller $\vperp$ values are expected for
         lensing of foreground disk sources since the source and lens would
         both share the Galactic rotation velocity of the Sun.
\end{enumerate}

Several authors who have modeled microlensing parallax events
\citep{mao-par,ogle2000bul43,ogle-par} have suggested that the source
stars must be predominantly in the foreground Galactic disk because this
makes a small $\vperp$ more likely. A disk source is the only possibility for
the OGLE-1999-CAR-1 event since this star is located far from the bulge,
but for events towards the Galactic bulge there are several factors that
make a foreground disk source star less likely, including a much lower
microlensing optical depth and a lower density of source stars. 
These are discussed in
section~\ref{sec-seleff}, where we find that disk sources that are
definitely in the foreground of the bulge at $D_s \leq 5\,$kpc are quite
unlikely.

\subsection{Projected Velocity Distributions}
\label{sec-vtdist}

One distinguishing characteristic of microlensing parallax distributions
for different source populations is the distribution of the projected
velocity, $\vpbold$ including both the amplitude, $\vp$, and the
direction $\theta$. We use a Galactic model in which
the stars around us are moving with a velocity
dispersion of about $30\,$km/sec in both directions normal to the line of
sight to the bulge. The Sun rotates at a speed of $+16\,$km/sec faster
than the {\it kinematic} Local Standard of Rest (LSR) and is moving
towards Galactic North at $7\,$km/sec \citep{lsr}. The Galactic disk
rotates with an approximately flat rotation curve at $v \simeq 200\,$km/sec,
while the Galactic bulge probably
has little rotation \citep{minniti96} and has a 
velocity dispersion of 80-$100\,$km/sec \citep{bul-pm,minniti96,hst-bul-pm}.
The Sagittarius Dwarf Galaxy is moving at $250\pm 90\,$km/sec in
a direction that is only a few degrees away from Galactic North
\citep{sgr-pm}.

\begin{figure}
\plottwo{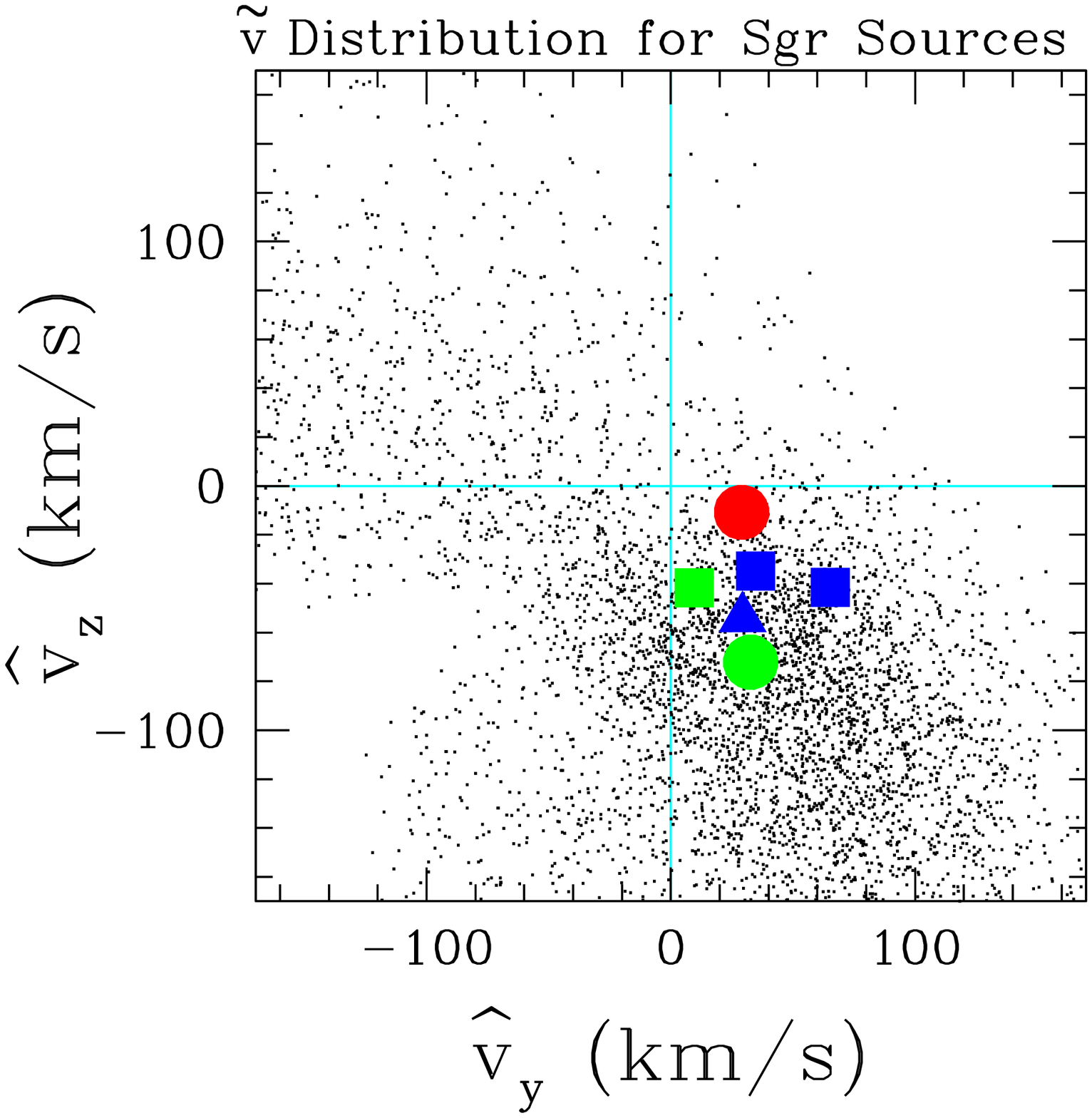}{blank.ps}
\caption{
The distribution of $\vpbold$ values in Galactic coordinates is 
shown for simulated microlensing parallax events towards the Galactic
bulge for three different source star populations: the Galactic bulge,
the foreground Galactic disk, and the Sagittarius Dwarf Galaxy.
The large colored dots show the locations of our detected microlensing
parallax events. The red circular spot is our best black hole candidate,
MACHO-96-BLG-5, and the green circular disk is the other black hole
candidate: MACHO-98-BLG-6. The two green spots are the events with
source stars that appear to be bulge sub-giants or Sagittarius Dwarf
red clump stars.  (MACHO-99-BLG-1 is the other). The blue squares are 
the bulge red clump source star events, and the
blue triangle is MACHO-98-BLG-8 which has a red giant source and is probably
also in the bulge.
\label{fig-vhatscat}}
\end{figure}

The different velocity distributions of these source and lens
populations lead to different expectations for the measured
$\vpbold$ distributions for events from different source star
populations. However, the observed $\vpbold$ distribution is
strongly affected by selection effects since only a small fraction
of microlensing events have detectable parallax signals. These
selection effects can be difficult to precisely quantify because of
the fact that much of the data taken for these events comes from
follow-up programs with observing strategies that can be subjective
and difficult to model.
Therefore, instead of attempting a detailed simulation of the actual
observing conditions, we investigate the $\vpbold$ distribution using
a ``toy model" of a microlensing survey and follow-up program. 
(\citep{kam-par-sim} also performed simulations of microlensing parallax
events in a somewhat different context.)
We assume a disk velocity dispersion of $30 \kms$ in each direction,
with a flat rotation curve of $200 \kms$ and a bulge
velocity dispersion of $80 \kms$ with no bulge rotation, and
the density profiles are a standard double-exponential disk and a barred
bulge as in \citet{hangould-mspec}.
We assume that events are observed for 7 months per year by a 
microlensing survey system that makes photometric observations with
5\% accuracy every 3 days. Once an event is magnified by at least
0.5 magnitudes, daily follow-up observations start with an accuracy
of 1\% for each day. This simulated data are then fit with a standard,
no-parallax microlensing model, and the $\Delta\chi^2$ is determined.
(Since we have not added noise to the light curves, the fit $\chi^2= 0$
when there is no microlensing parallax signal.)
Events with $\Delta\chi^2 \geq 200$ are considered microlensing parallax
detections, and the $\vpbold$ values for these simulated detected
events are shown in Figure~\ref{fig-vhatscat}. This figure uses
Galactic coordinates in which the $y$-axis is the direction of Galactic
disk rotation, and the $z$-axis is Galactic North.

A striking feature of Figure~\ref{fig-vhatscat} is that all six of our
strong microlensing parallax events have $\vpbold$ in the same quadrant
with positive $\vp_y$ and negative $\vp_z$. This is the region that
is preferred for both bulge and SGR source stars, but not for foreground
disk sources. In our simulations, 65\% of the detectable SGR source events,
and 50\% of the detectable bulge sources, but only 29\% of foreground
disk sources lie in this quadrant.

One selection effect that affects each plot is that events with $\vpbold$
roughly parallel to the ecliptic plane are easier to detect than events
where $\vpbold$ is approximately perpendicular to the ecliptic plane.
This effect favors the positive $\vp_y$-negative $\vp_z$ and
negative $\vp_y$-positive $\vp_z$ quadrants. The reason for this is that
the Earth's orbital motion only affects $u(t)$ near peak magnification
when $\vpbold$ is perpendicular to the ecliptic plane, but the orbital
motion affects $u(t)$ for a longer period of time when it is parallel
to $\vpbold$.

For the bulge sources, there is a preference for positive $\vp_y$ motion
because disk lens stars are passing inside of us at a higher angular velocity.
If the source stars are rotating with us, as would be the case for
disk sources in the foreground of the bulge, then the rotation is
common to the source, lens and observer, and it has no effect. A smaller
systematic effect occurs in the disk source case because the Sun is moving
about $16\,$km/sec faster than the mean stellar motion around us. Thus, there
is a slight enhancement of the abundance of negative $\vp_y$ events.

For SGR source stars, signal of the SGR proper motion toward the
Galactic North can clearly be seen in the strong concentration of events
at negative $\vp_z$ and positive $\vp_y$. (Since $\vpbold$ is a
lens$-$source velocity, the $\vpbold$ signal is in the opposite direction of
the SGR motion.) For bulge lenses, the $250\,$km/sec velocity is reduced to
$90\,$km/sec by the projection effect, and for the disk lenses that make up
the bulk of microlensing parallax sample for SGR sources,
the typical $\vp_z$ is $-50\,$km/sec or so. For SGR sources,
and disk lenses, the combination of SGR proper motion and disk rotation
put the majority of $\vpbold$ values in the positive $\vp_y$-negative $\vp_z$
quadrant where the alignment with the ecliptic plane makes the parallax
effect easy to detect.


\subsection{Microlensing Parallax Selection Effects}
\label{sec-seleff}

For comparison between the different source populations, it is
necessary to consider several different selection effects. First, the
microlensing rate for SGR sources behind our fields is a 
factor of $\sim 6$ larger than for bulge source stars \citep{sgr-lensing}, 
and the fraction of SGR events with
detectable microlensing parallax signals is a factor of 3 larger for
sources in SGR than for bulge sources. Thus, it would appear that the
probability of detecting a microlensing parallax event for a
SGR source is a factor of $\sim 20$ higher than for a bulge source
(assuming that the sources are bright enough for reasonably accurate 
photometry). Of course, Galactic bulge sources are much more numerous,
so microlensing parallax events with Galactic bulge source stars are 
likely to be more numerous than events with SGR source stars
by an amount that is difficult to estimate. We consider this in detail
in Section~\ref{sec-cmd} when we present the source star color-magnitude
diagrams.

It is quite difficult to distinguish Galactic bulge stars from stars
in the inner Galactic disk because they are at similar distances
and their velocity distributions overlap.
In fact, this distinction is likely to be somewhat artificial because
the two components are likely to have merged due to their
mutual gravitational interactions. Therefore, we will limit
our consideration of foreground disk sources to stars with a distance
$< 5\,$kpc. For stars at $5\,$kpc distance at a Galactic latitude of
$b = -3^\circ$, the microlensing rate is a factor
of about $\sim 40$ lower than for Galactic bulge stars. (The optical depth is
only a factor of $\sim 20$ lower because of the longer time scales of disk-disk
lensing events.) The physical density of disk stars is about an
order of magnitude lower than the density of bulge stars, but there
is also a volume factor that reduces the number of disk stars per unit
distance modulus and solid angle by a factor of 4 at $5\,$kpc. The product of 
these factors yields a net suppression factor of $1/1600$ for disk star
lensing events for a fixed source star absolute magnitude.

This suppression factor must be multiplied by two enhancement factors.
First, our simulations indicate that the chances of detecting
a microlensing parallax signal are about a factor of 5 larger for
disk sources than for bulge source stars. This increases the suppression factor
to $\sim 1/320$. There is an additional enhancement factor due to
the fact that the foreground disk stars are intrinsically fainter and
the stellar luminosity function rises for fainter stars, but the
difference between the disk stars at $5\,$kpc and the bulge stars
at $8\,$kpc is only 1 magnitude. From \citet{holtzman-bul} we see that
this factor is at most $\approx 10$ if we select a source magnitude such
that is 1-2 magnitudes above the bulge main sequence turnoff. For
magnitudes that correspond to bulge main sequence stars, it is less than
a factor of two. Thus, we expect disk stars (with $D < 5\,$kpc)
to contribute less than 1\% of the total
number of detectable microlensing parallax events,
except for source stars that are 1-2 magnitudes above the bulge main sequence
turn-off where they might account for as many as 3\% of the
microlensing parallax events with bulge source stars.

Inner disk stars at $D>5\,$kpc will be accounted for by allowing their
velocities to contribute to the assumed bulge velocity distribution.
In fact, such inner disk stars are generally not
excluded from star samples that are used to measure the bulge proper motion
\citep{bul-pm,hst-bul-pm}. We will, therefore, classify all stars in
the vicinity of the bulge ($5\,{\rm kpc} < D_s < 11\,{\rm kpc}$) as bulge
stars. Instead of trying to distinguish different, but overlapping,
populations of source stars, we consider a single model including
all these stars.

Stars on the far side of the disk have velocities that
make it very unlikely to see the microlensing parallax effect,
while foreground disk stars are unlikely to be microlensed at all.
Therefore, the SGR dwarf provides the only ``non-bulge" population
of potential source stars that we will consider in the remainder of this
paper. 

\subsection{Color Magnitude Diagrams}
\label{sec-cmd}

\begin{figure}
\plottwo{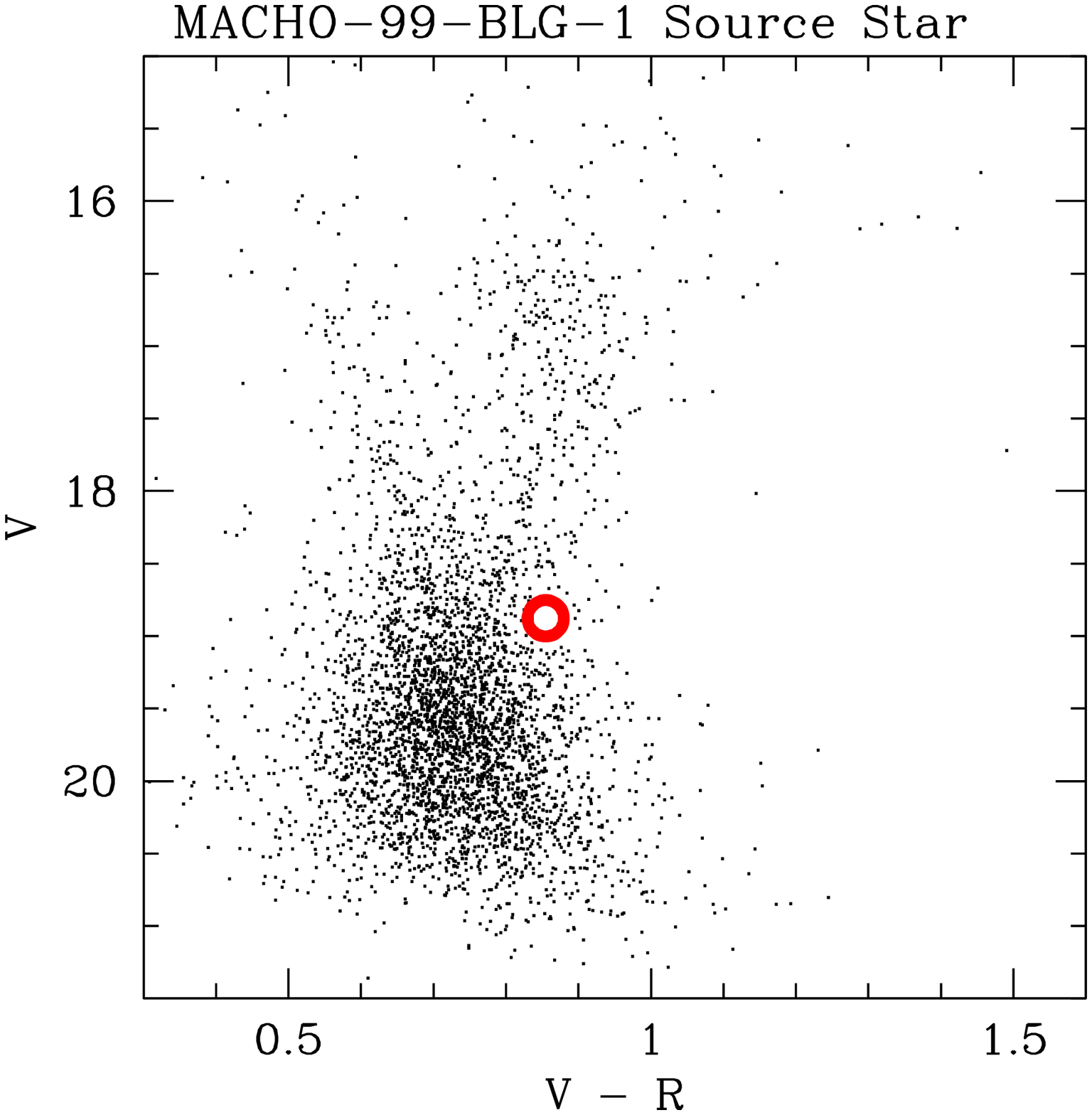}{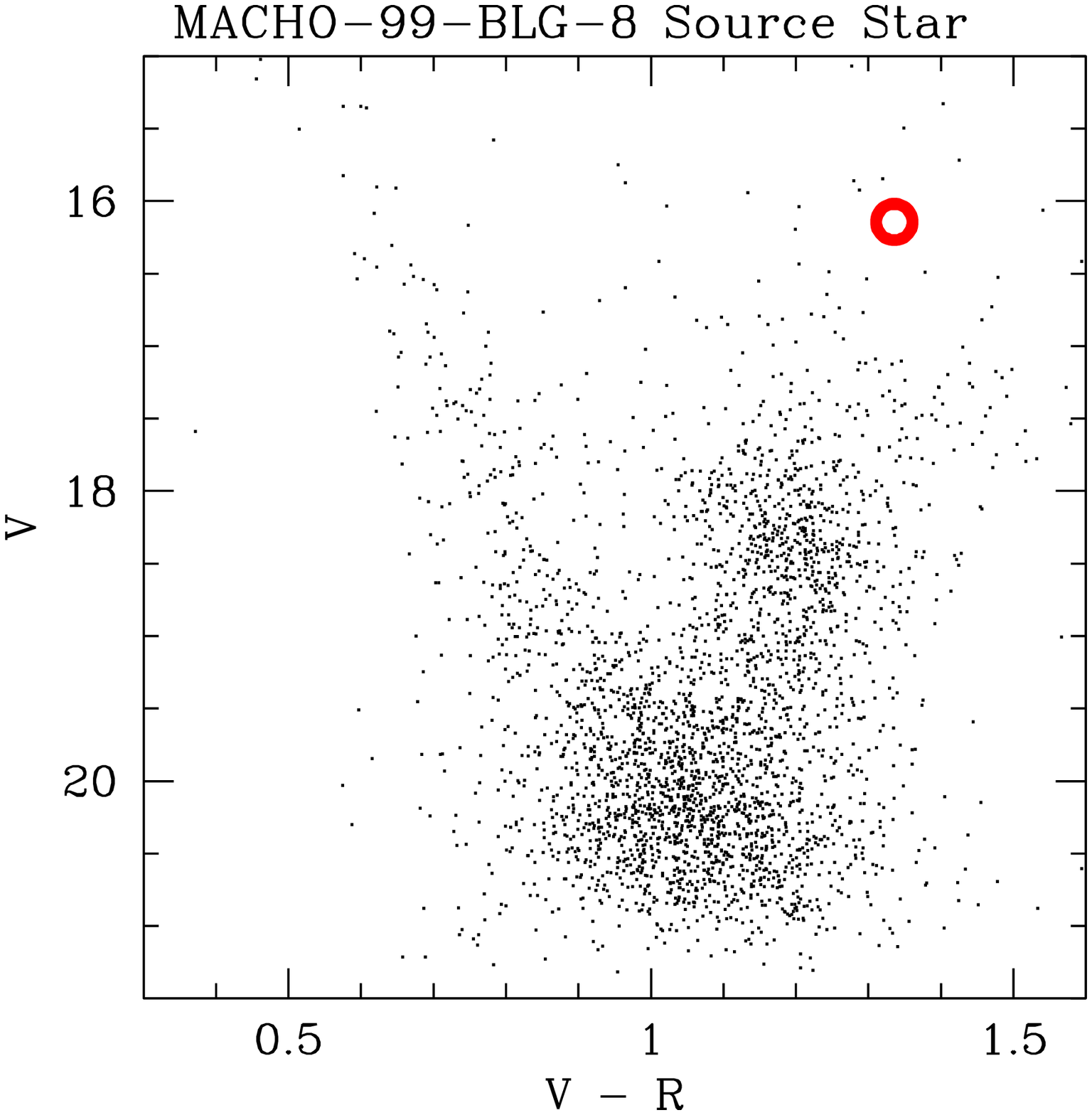}
\caption{
Color-Magnitude diagrams from MACHO data are shown for all the detected
stars within a 2 arc minute circle around each of our 
microlensing parallax events. The red circles indicate the location of
the lensed source star, assuming the blending implied by the microlensing
parallax fits.
\label{fig-cmds}}
\end{figure}

Fig.~\ref{fig-cmds} shows color-magnitude diagrams for all the stars within
2 arc minutes around each of our microlensing parallax
source stars, with the lensed source 
indicated by a red circle. It is necessary to use different color magnitude
diagrams for each event because of the large variation in reddening between
different fields. By plotting only the stars within 2 arc minutes of our
targets, we have minimized the variation in reddening.

These CM diagrams indicate that the MACHO-104-C and MACHO-96-BLG-12 source 
stars are located in the bulge red clump region, 
which means that they are likely to reside in
the Galactic bulge. The MACHO-99-BLG-8 source star is more luminous than
the red clump and is likely to be a bulge giant. \citet{cook-spectra} find
a radial velocity of $v_r = 195\pm 2 \kms$ which confirms the bulge
interpretation for this event.
The MACHO-96-BLG-5 source star appears to be fainter
than virtually all of the other stars in its CM diagram. This is a consequence
of the extreme crowding of these Galactic bulge fields. The density of bright
main sequence stars is $\sim 2$ per square arc second, so main sequence
stars are not individually resolved in these crowded Galactic fields.
Instead, it is groups of unresolved main sequence stars that are identified
as single stars, and it is these unresolved blends of multiple stars
that make up the majority of the fainter objects identified as stars in these
images. The majority of microlensed source stars in the Galactic bulge
are blended main sequence stars like the MACHO-96-BLG-5 source, but the
microlensing parallax signal is easier to detect for brighter source stars.

The source stars for events MACHO-98-BLG-6 and MACHO-99-BLG-1 appear to
be on the bulge sub-giant branch of the color magnitude diagram. They
have a similar color to bulge red clump stars, but they are about 2 magnitudes
fainter. This suggests that they could be red clump stars $\sim 14\,$kpc
behind the bulge in Sagittarius (SGR) Dwarf Galaxy. This is about the only
location on the color magnitude diagram were we might expect to see
microlensing of SGR source stars, because SGR red clump stars are
probably the only abundant type of SGR stars that are brighter than
the bulge main sequence stars that set the confusion limit.
This SGR source interpretation appears to gain support from the 
location of these events in Fig.~\ref{fig-vhatscat} which 
indicates that their parallax velocities are among the ones most 
consistent with Sagittarius Dwarf kinematics.

A rough estimate of the probability of detecting microlensed
SGR source stars can be made by noting that SGR Dwarf RR Lyrae stars are
about 2.6\% as numerous as bulge RR Lyrae in the MACHO
fields \citep{macho-sgr-rrl}. In a microlensing parallax sample,
we should expect SGR source stars to be enhanced by a factor of
$\sim 20$, but we must also include both bulge sub-giants and
giants in the comparison with the SGR red clump giants. This
would reduce the fraction of SGR events by a factor two or so.
This would suggest that we might expect that for every 4
microlensing parallax events with bulge giant or sub-giant source
we could expect one SGR giant source star event\rlap.\footnote[2]
{A previous estimate of lensing rates for SGR source stars has
been made by \citet{sgr-lensing} who find a smaller ratio of SGR/bulge 
source lensing events 
than our estimate. This is because they
do not consider only microlensing parallax events and because they
count the much more numerous events with bulge main sequence source
stars.} On the other hand, the ratio of red clump stars to RR Lyrae
is likely to be higher for SGR stars than for Galactic bulge stars
because of the lower metalicity of SGR, so we might expect fewer
SGR events than this RR Lyrae comparison would suggest.

These considerations suggest that we should take the SGR source star
hypothesis seriously for these events. However, \cite{cook-spectra}
used the Keck HIRES spectrograph to obtain spectra of the source stars
for these events, and they find radial velocities of $v_r = -65\pm 2\kms$
and $v_r = 64\pm 2\kms$ for MACHO-98-BLG-6 and MACHO-99-BLG-1,
respectively. This is not consistent with the SGR radial velocity
\citep{sgr-pm} of $v_r = 140\pm 10\kms$, and they are about $2\sigma$
away from the expectation for a disk source star
\citep{disk-vr}. Thus, these events are most likely to have
bulge sub-giant source stars.

\subsection{Likelihood Distance and Mass Estimates}
\label{sec-likeli}

Another, somewhat more general, constraint on $x$ and $M$ can be 
obtained if we make use of our knowledge of the velocity 
distributions of the source and lensing objects, 
since the likelihood of obtaining the observed 
value of $\vpbold$  is a strong function of the distance
to the lens. Note that this assumes that stellar remnant lenses have
a velocity and density distribution that is similar to that of observed
stellar populations. For neutron stars, this might be a questionable
assumption because many neutron stars are apparently born with a large
``kick" velocity. However, for black holes, the evidence indicates
that significant kick velocities are rare \citep{bh-kick}.
As an example of such an analysis, let us suppose that the disk and 
bulge velocity dispersions
were negligible relative to the Galactic rotation velocity.
Then, for disk lenses we would obtain the relation 
$\vp = 200 D_\ell / (D_s - D_\ell ) \kms$ implying a lens 
distance of $D_{\ell} = D_s \vp/(\vp + 200 \kms)$.
In reality, the random motions of both disk and bulge stars 
broaden this relationship somewhat, 
but we can still obtain a useful constraint. 

Given the observed $\vpbold$, we obtain a likelihood function
\begin{equation}
 L(x;\vpbold) \propto \sqrt{x (1-x)} \, \rho_L (x) \, \vp (1-x)^3 
\int f_S(\vs) \, f_L( (1-x) ({\bf v}_\odot + \vpbold) + x \vs ) \, d\vs ,
  \label{eq-like}
\end{equation}
where $\rho_L$ is the density of lenses at distance $x = D_\ell/D_s$,
and the integral is over combinations of source and lens velocities
giving the observed $\vpbold$. $\vs$ and
$\vl=(1-x) ({\bf v}_\odot + \vpbold) + x \vs$ are 
the 2-D source and lens velocity distribution functions (normalized to unity).
We assume the same Galactic parameters as in our $\vp$ simulations above:
a disk velocity dispersion of $30 \kms$ in each direction, 
a flat disk rotation curve of $200 \kms$, and a bulge 
velocity dispersion of $80 \kms$ with no bulge rotation. 
The density profiles are a standard double-exponential disk and a 
\citet{hangould-mspec} barred bulge.
For all events, the source is assumed to reside in the bulge, 
while the lens may be in the disk or the bulge. But for events MACHO-98-BLG-6
and MACHO-99-BLG-1, we also consider the possibility of a SGR Dwarf 
source star with the lens in the disk or bulge, although this now
appears to be ruled out \citep{cook-spectra}.

\begin{figure}
\plottwo{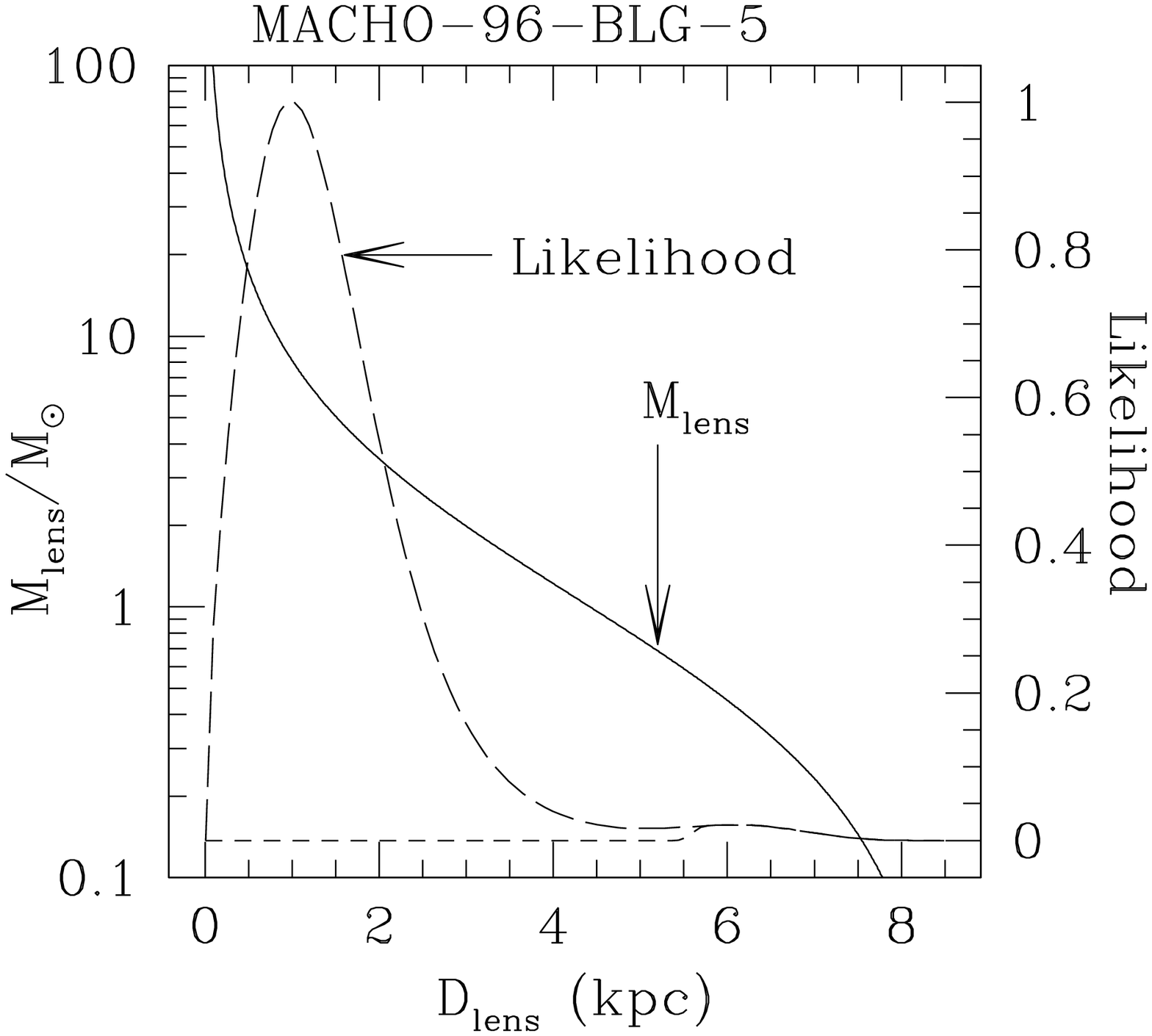}{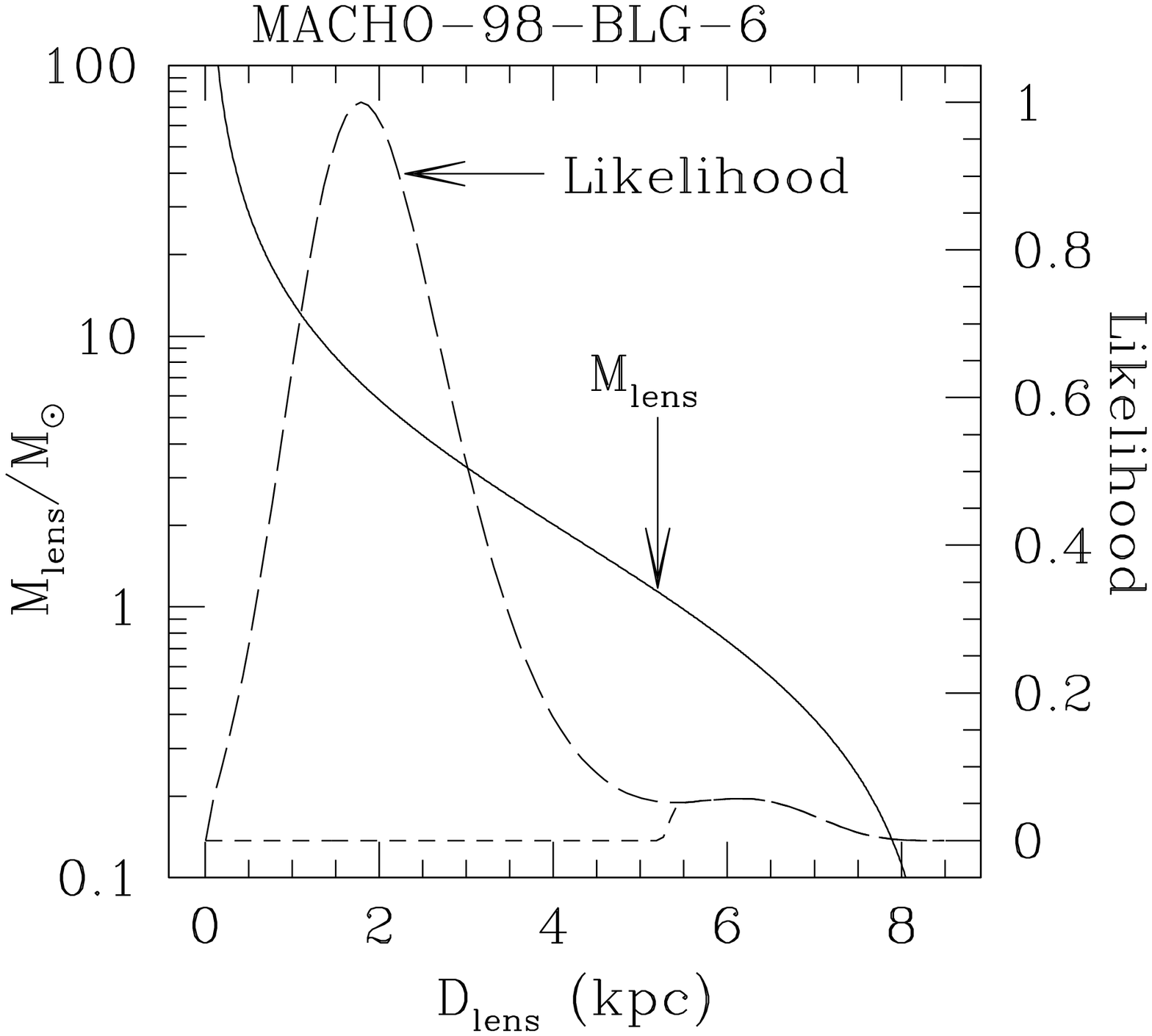}
\caption{
The mass vs.~distance relations (solid curves) for our candidate black hole 
lenses are shown along with the likelihood functions (long dashed curves)
computed assuming a standard model for the Galactic phase space distribution.
The source star is assumed to reside in the bulge for both events. The implied
best fit masses are $M = 6{+10\atop -3}\msun$ for the MACHO-96-BLG-5 lens
and $M = 6{+7\atop -3}\msun$ for the MACHO-98-BLG-6. The 95\% confidence
level lower limits on the masses are $1.6\msun$ and $0.94\msun$ respectively.
The short dashed curves delineate the portion of the likelihood functions 
that is allowed when the lens is assumed to be a main sequence
star. The ratio of the area below this portion to the entire area below the
likelihood curve gives a probability that a lens is a main sequence star.
For MACHO-96-BLG-5, the
upper limit on the lens brightness is very stringent because of the HST
images, and a main sequence lens is ruled out.
\label{fig-masslike-bh}}
\end{figure}

\begin{figure}
\plottwo{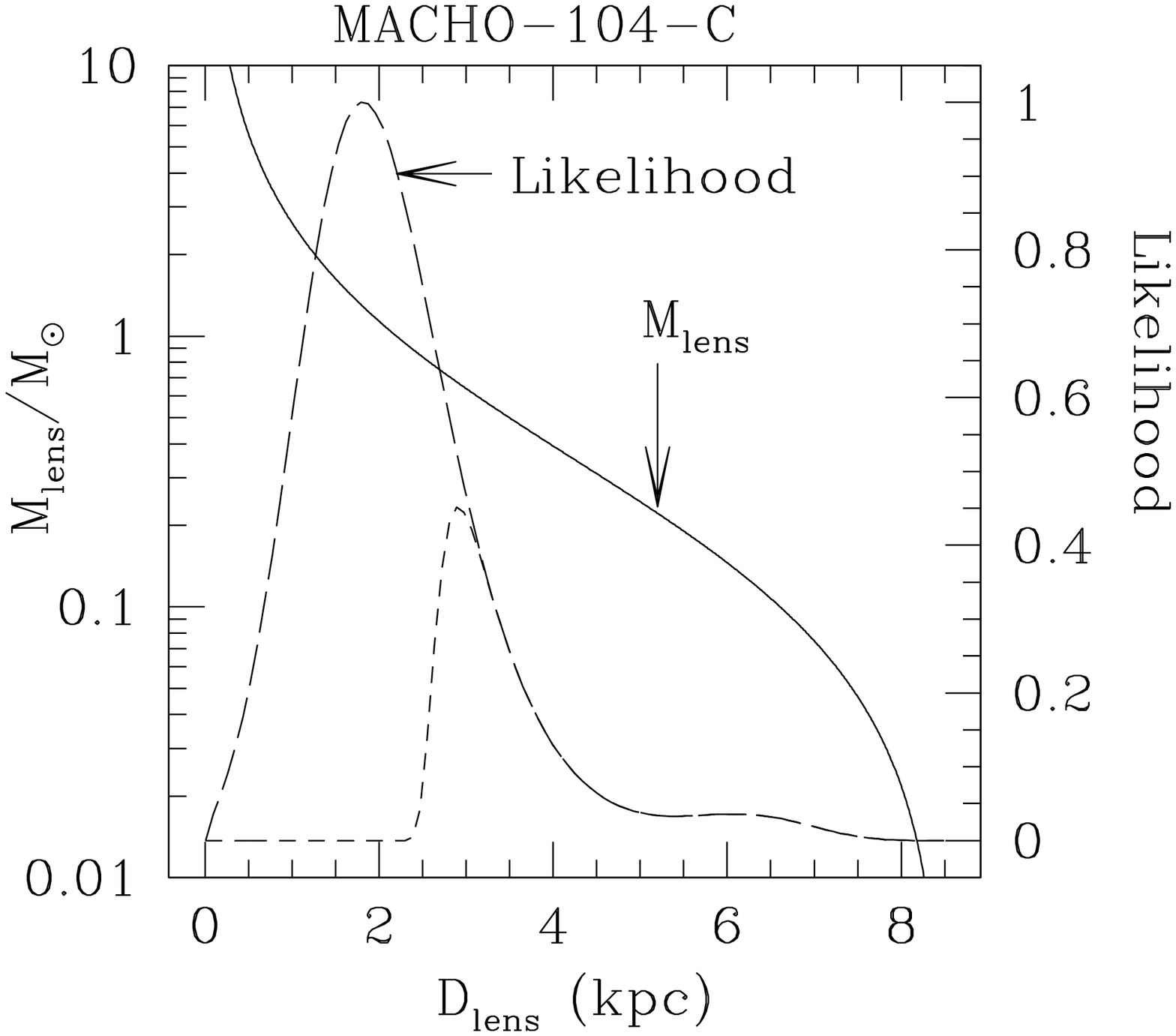}{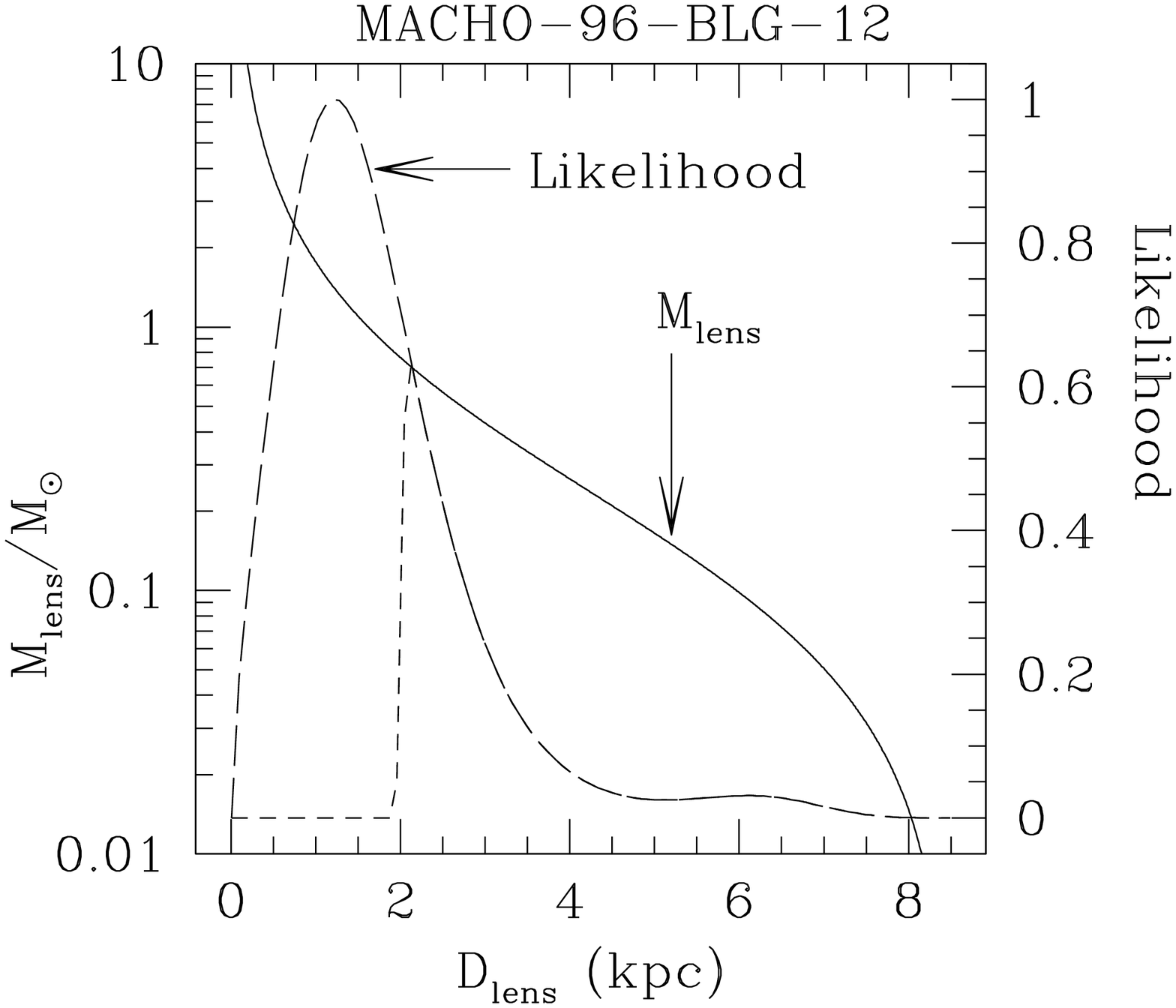}
\caption{
The mass vs.~distance relations (solid curves) for our two bulge clump 
giant source events are shown along with the likelihood functions 
(long dashed curves) computed assuming a
standard model for the Galactic phase space distribution. The implied
best fit masses are $M = 1.1{+1.1\atop -0.5}\msun$ for the MACHO-104-C lens
and $M = 1.3{+1.8\atop -0.7}\msun$ for the MACHO-96-BLG-12. The 95\% confidence
level lower limits on the masses are $0.35\msun$ and $0.33\msun$ respectively.
The short dashed curves delineate the portion of the likelihood functions 
that is allowed when the lens is assumed to be a main sequence
star, and they indicate that
main sequence lenses are disfavored but not ruled out.
\label{fig-masslike-clump}}
\end{figure}

\begin{figure}
\plottwo{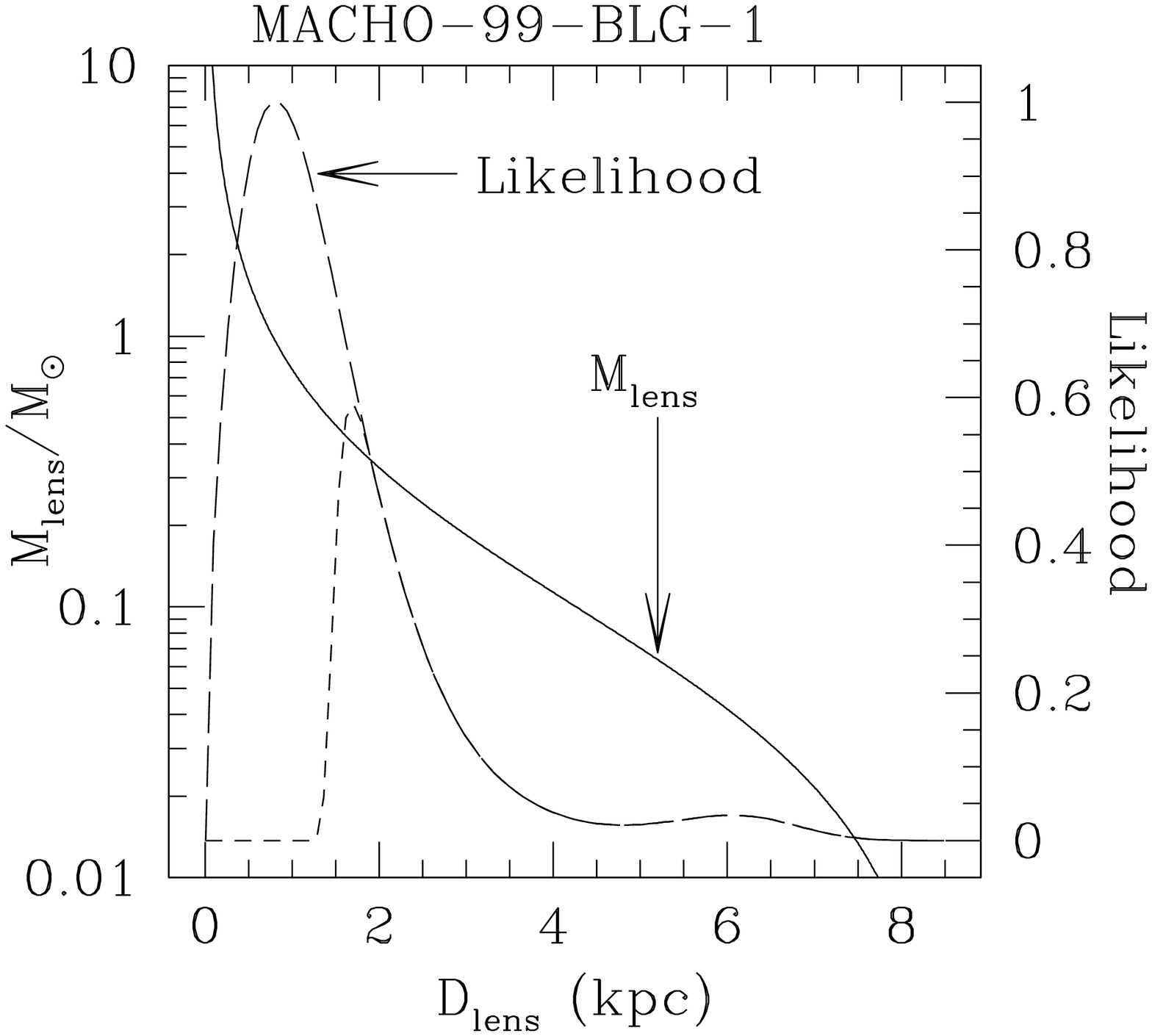}{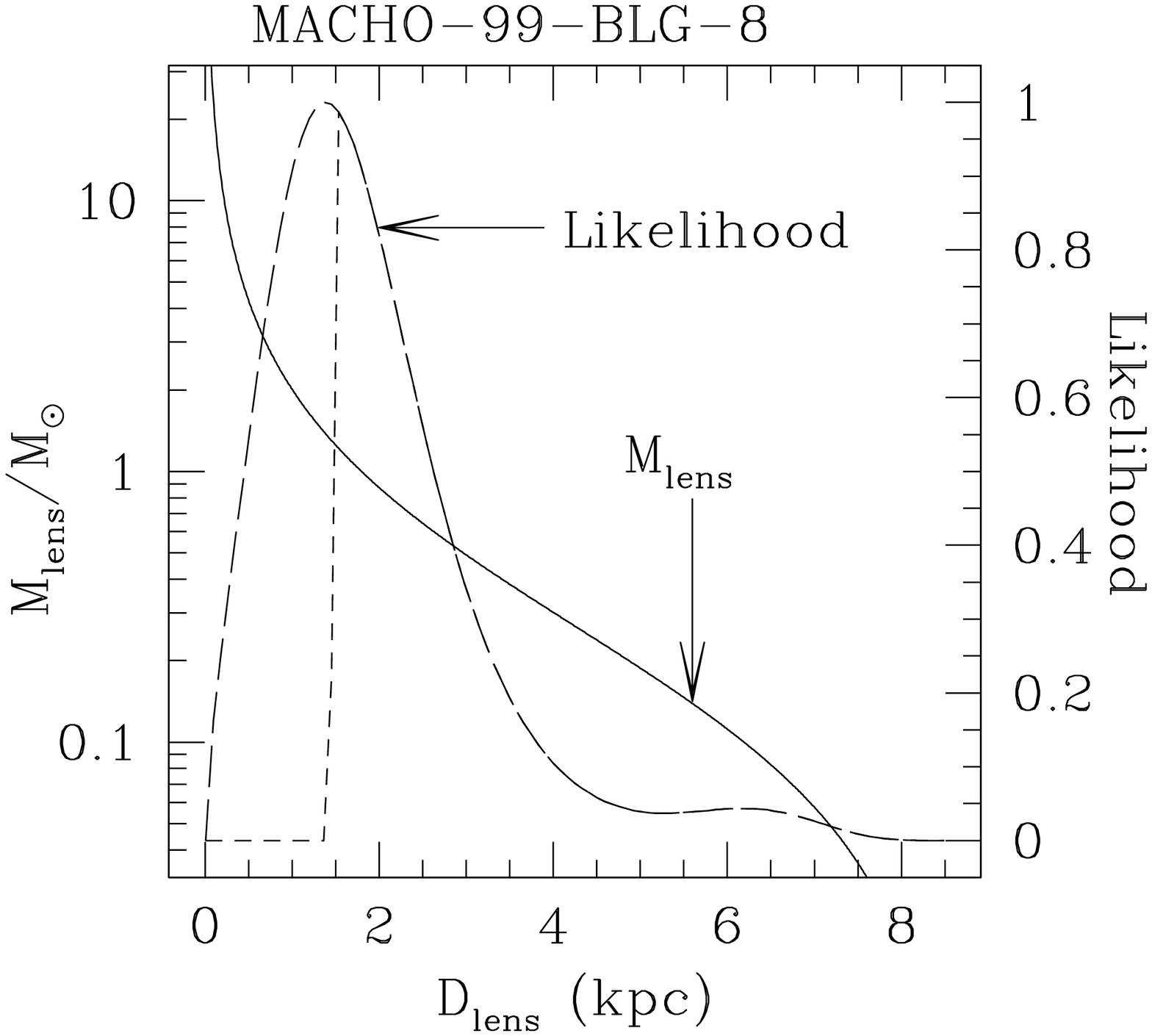}
\caption{
The mass vs.~distance relations (solid curves) for the two 1999 microlensing 
parallax events are shown along with the likelihood 
functions (long dashed curves) computed assuming a
standard model for the Galactic phase space distribution. For both events
the source star is assumed to reside in the Galactic bulge.  The implied
best fit masses are $M = 0.7{+1.2\atop -0.4}\msun$ for the MACHO-99-BLG-1 lens
and $M = 1.2{+1.6\atop -0.6}\msun$ for the MACHO-99-BLG-8. The 95\% confidence
level lower limits on the masses are $0.14\msun$ and $0.3\msun$ respectively.
The short dashed curves delineate the portion of the likelihood functions 
that is allowed when the lens is assumed to be a main sequence
star, and these indicate that
the lens brightness constraints are consistent with main sequence
lens stars. For MACHO-99-BLG-1, a main sequence lens is disfavored, however.
\label{fig-masslike-99}}
\end{figure}

\begin{figure}
\plotone{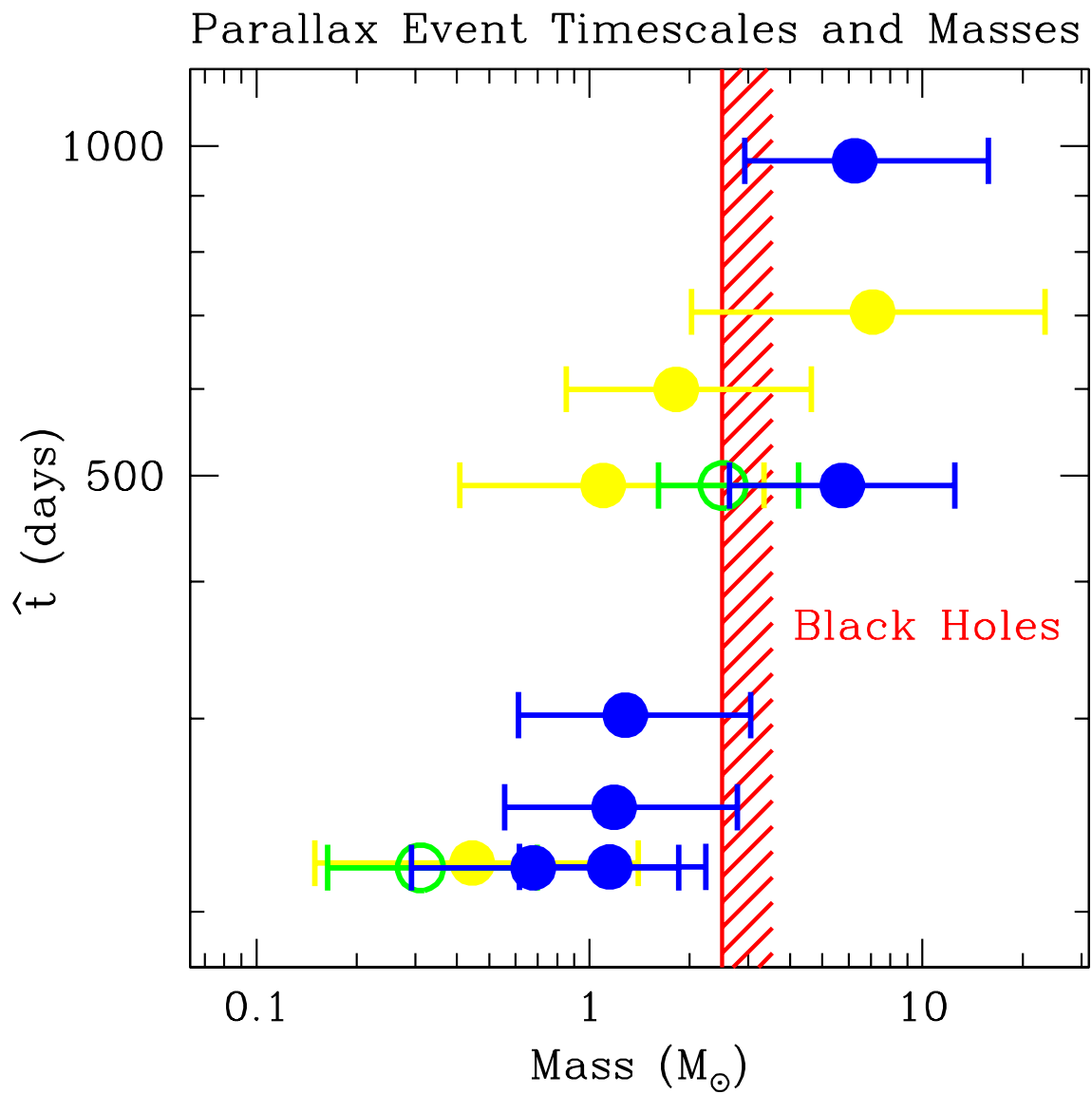}
\caption{
This plot shows $M$ vs.~$\that$ for the 10 events with $\that > 140\,$ days 
with $1\sigma$ error bars for the mass estimates. All events with
formally significant detections of microlensing parallax are
shown, but
parallax signal for the events indicated in yellow is weak enough that
the detection is not considered to be definitive. The green open symbols
indicate the predicted lens masses for MACHO-98-BLG-6 and MACHO-99-BLG-1
if their source stars were in the SGR Dwarf Galaxy, a possibility that 
appears to be contradicted by their spectra.
\label{fig-m_vs_that}}
\end{figure}

The resulting likelihood functions for 
$D_{\ell}$ is shown as the long-dashed curves in 
Figs.~\ref{fig-masslike-bh}-\ref{fig-masslike-99},
and these are insensitive to specific parameter choices. These likelihood
functions also provide a means for estimating the lens masses
via the relation~(\ref{eq-m}), which is also plotted in 
Figs.~\ref{fig-masslike-bh}-\ref{fig-masslike-99}. 
Fig.~\ref{fig-m_vs_that} shows how
the mass estimates correlate with the best fit event timescale for the
six high signal-to-noise microlensing parallax events as well as four
other events of lower signal-to-noise. (The lower signal-to-noise event
with the highest mass is MACHO-99-BLG-22/OGLE-1999-BUL-32 which has
been presented as a black hole candidate by Mao et al.~2001.)

One common way to interpret likelihood functions is the Bayesian method,
in which the lens mass (or distance) probability distribution is given
by the likelihood function times a prior distribution, which represents
our prior knowledge of the probability distribution. In our case, the
likelihood function represents all of our knowledge about the lens mass
and location, so we select a uniform prior. With a uniform prior, the
likelihood function becomes the probability distribution and
we are able to 
calculate the lens mass confidence levels listed in Table~\ref{tbl-mass}.
This table also includes lens mass confidence levels for models
that differ from the preferred model in order to show how the mass
estimates depend upon the amount of blending (for MACHO-96-BLG-5) and
on whether the source star resides in the Galactic bulge or the
SGR Dwarf. Note that the uncertainty in the mass estimates is smaller for
SGR Dwarf sources due to the small velocity dispersion of the SGR Dwarf
and the smaller range of likely lens distances.

\subsection{Constraints on Main Sequence Lenses}
\label{sec-mslenses}

If we assume that the lens stars are main sequence stars, then
we can obtain an additional constraint on their distances and masses
by comparing the brightness of a main sequence star, of the implied
mass, to the upper limit
on the brightness of the lens star.  We have  assigned
a conservative upper limit on the V-band brightness of each lens star
based upon the available photometry and microlensing parallax fits
listed in Table~\ref{tbl-fitpar}. In the case of MACHO-96-BLG-5, the
upper limit is particularly stringent because it is based upon HST
observations. Note that if we assign some of the flux of the star
identified in the HST images to the lens star instead of the source,
the best fit $\that$ will increase almost linearly with the inverse of
the source star flux. This causes the lens mass estimate to increase
as $\sim \that^2$. Since stellar luminosity varies as a high power
of the mass, a main sequence lens will be more strongly ruled out.

In order to apply these constraints to the likelihood
functions for the mass and distances of the lens stars,
we have multiplied the likelihood function by the Gaussian probability that
the lens brightness exceeds the upper limit on the brightness of
lens star. If a main sequence lens star would be fainter than 
the observed maximum brightness, there is no modification of
the likelihood function. 
This gives the short-dashed likelihood curves shown in 
Figs.~\ref{fig-masslike-bh}-\ref{fig-masslike-99}.
These results are insensitive to our assumed $L \propto M^4$
mass-luminosity relation. The assumed maximum lens brightnesses are 
$V = 19.88$, $20.57$, and $16.92$ for events MACHO-96-BLG-12, MACHO-98-BLG-6,
and MACHO-99-BLG-8, respectively. These are based upon the amount of
blending allowed by the fit, and each of these has an assumed 25\% 
uncertainty which is also based upon the fit. For MACHO-96-BLG-5, the
maximum lens brightness is $V = 23.63$, with an assumed 50\% uncertainty.
For MACHO-104-C, and MACHO-99-BLG-8, the best fit has very little
blended flux: $V = 22.33$, and $23.08$, respectively. But, in both cases,
the uncertainty in the blended flux is five times the best fit value.


The properties of the most likely main sequence
lens models are given in Table~\ref{tbl-mmag}, which is discussed in
more detail in section \ref{sec-lensdetect}. An important parameter in
this table is the predicted lens-source separation in June, 2003, when they
might plausibly be observed by HST. This
can be calculated from the lens-source proper motion which is 
related to the projected velocity by $\mu = \vp (D_s - D_\ell)/(D_s D_\ell)$.

\subsection{Stellar Remnant Lenses and Black Hole Candidates}
\label{sec-remnant}

The mean mass estimate for the six microlensing parallax events is $2.7\msun$.
Five of the six have best fit masses $> 1\,\msun$, and two of the
events, MACHO-96-BLG-5 and MACHO-98-BLG-6, have best fit
masses $> 3\msun$. This makes them black hole candidates because the
maximum neutron star mass is thought to be $\sim 2\msun$ \citep{ns-max-mass}.
The 95\% confidence level lower limits on the masses of these lenses are
$1.64\msun$ and $0.94\msun$, respectively, while the 90\% confidence level 
lower limits are $2.3\msun$ and $1.9\msun$. A main sequence star lens
at the lower limit mass is strongly excluded in the case of MACHO-96-BLG-5
because of the constraint on the lens brightness from HST images.
However, a main sequence lens with a mass at the 95\% confidence limit
is not quite excluded for MACHO-98-BLG-6.
The masses that have been measured for neutron stars
are close to the Chandrasekhar mass,
$M_{NS} = 1.35\pm 0.04 \msun$ \citep{ns-mass}, which is excluded at
better than 95\% confidence for MACHO-96-BLG-5 and better than 90\%
confidence for MACHO-98-BLG-6. Thus, both MACHO-96-BLG-5
and MACHO-98-BLG-6 are both black hole candidates, but there is a small
chance that MACHO-98-BLG-6
could be a neutron star or even a main sequence star.

In addition to these black hole candidates, three of the remaining four
microlensing parallax events have best fit masses $> 1\msun$. For
MACHO-104-C and MACHO-96-BLG-12, main sequence lens are disfavored,
but not ruled out. MACHO-99-BLG-8 appears to be blended with a relatively
bright source, so a main sequence lens of $M \simgt 1\msun$ is a
possibility. As we explain below, with HST imaging it will be
straightforward to detect the lenses if they are main sequence stars.
If HST images fail to detect the lens stars, then we can show that
the lenses are almost certainly stellar remnants.

\subsection{Likelihood Analysis with a Mass Function Prior}
\label{sec-mprior}

The Likelihood analysis presented in Section~\ref{sec-likeli} attempts
to estimate the distance to the lens based upon the measured value of the
projected velocity, $\vpbold$, and then the lens mass is determined
from eq.~\ref{eq-m}. If the lens mass function, $dn/dM = \phi(M)$,
is known, then it is possible to use the measured $\that$ value
to make a more accurate estimate of the lens mass as advocated by
\citet{agol}. The likelihood function,
eq.~\ref{eq-like}, can be modified by multiplying by 
$\delta(\that-\that_m) M^{1/2} \phi(M) dM$ and integrating over
$M$, where $\that_m$ is the measured value of $\that$. The factor of
$M^{1/2}$ is the contribution of the lens mass to the lensing
cross section, which is proportional to $R_E$. The integral over 
$\delta(\that-\that_m) dM$ gives an additional factor of $M$.
Thus, the likelihood analysis presented in Section~\ref{sec-likeli}
is equivalent to assuming a mass function of $\phi(M) \propto M^{-1.5}$.

A more conventional mass function for the Galactic bulge is a broken power law
initial mass function \citep{kroupa} with $\phi(M) \propto M^{-1.3}$ for 
$0.03{\msun} \leq M \leq 0.8\msun$, and $\phi(M) \propto M^{-2.35}$ for
$0.8{\msun} \leq M \leq 100 \msun$. However, the stars with $M > 1.0\msun$
will generally have ended their main sequence lifetimes and have become
stellar remnants after significant mass loss. Following \citet{fryer01},
we can assume that all stars with an initial mass 
greater than a particular cutoff mass, $M_i > M_{\rm BH}$ become black holes.
We take $M_{\rm BH} = 20\msun$ \citep{fryer99,fryer01}. 
Similarly, we assume that all stars with 
$8{\msun} \leq M_i < M_{\rm BH}$ become neutron stars, and all stars with 
$ 1.0{\msun} < M_i < 8{\msun}$ become white dwarfs. The mass functions of
the stellar remnants are assumed to be Gaussians with mean masses of
$0.6\msun$, $1.35\msun$, and $8\msun$ for white dwarfs, neutron stars,
and black holes respectively. The Gaussian sigmas are $0.15\msun$, 
$0.04\msun$, and $2.5\msun$, respectively. These are consistent with the
measured mass functions \citep{wd-mass,wd-cool-mass,ns-mass,bh-mass}, 
although the difficulty of directly observing old stellar remnants 
assures that the observed samples are incomplete. With this mass function,
black holes would account for 3.7\% of the Galaxy's stellar mass.

A Bayesian analysis based upon this mass function gives a probability
of 93\% that the MACHO-96-BLG-5 lens is a black hole and a probability of
69\% that the MACHO-98-BLG-6 lens is a black hole. The probability of at
least one black lens is 98\%. This analysis may underestimate 
the black hole probability because the assumed mass function cannot account
for the large number of long timescale microlensing events. 
An initial IMF that is slightly shallower than
the Salpeter slope, $\phi(M) \propto M^{-2.0}$, might be appropriate
if most of the stars in the Galaxy were formed in denser or more metal
poor regions than is typical for present day star forming regions
\citep{arches,m82f}.  With this mass function and with $M_{\rm BH} = 20\msun$, 
black holes would account for 12\% of the Galaxy's stellar mass. When we 
repeat the likelihood analysis with
this mass function, we find black hole probabilities of 97\% for 
MACHO-96-BLG-5 and 88\% for MACHO-98-BLG-6. The probability of at least
one black hole lens with this mass function is 99.7\%. If we retain
the Salpeter IMF slope, and increase $M_{\rm BH}$ to $40\msun$, then
the black hole probabilities for MACHO-96-BLG-5 and MACHO-98-BLG-6
drop to 82\% and 43\%, respectively. However, such a mass function probably
cannot explain the excess of long timescale events.

We should note
that these probabilities are substantially larger than those reported
in a similar analysis in a preprint by \citet{agol}. This was
due to a likelihood function calculation error by \citet{agol}. When this
error is corrected, their results are quite similar to those presented
here (Agol, private communication).  

\section{Follow-up Observations}
\label{sec-followup}

The detection of the microlensing parallax effect allows us
to make a lens mass estimate that is accurate to about a factor of two,
and to identify the black hole candidates. However, these estimates
are not accurate enough to determine the black hole mass function,
and they do not allow the unambiguous identification of neutron
star or white dwarf lenses. However, follow-up observations with
higher resolution instruments hold the promise of much more
precise determinations of the lens masses.

\subsection{Interferometric Follow-up}
\label{sec-interfere}

The most ambitious of microlensing event follow-up plans involve
interferometric instruments such as the Keck and VLT interferometers 
\citep{vlti} and the Space Interferometry Mission (SIM)
\citep{boden}. The most spectacular confirmation of a black hole
event would be to measure the image splitting which is given by
\begin{equation}
\phi_{\rm sep} = 2\theta_E \sqrt{1+u^2/4} \ ,
\label{eq-split}
\end{equation} 
where $\theta_E$ is the image separation and $u$ is given by
eq.~(\ref{eq-upar}). For MACHO-96-BLG-5,
we have $\theta_E = 9.8\,$mas if the lens is at the distance
preferred by the likelihood analysis. This compares to the
$5\,$mas diffraction limit of an interferometer with a $100\,$m
baseline operating at a wavelength of $2\,\mu$m, such as the 
Keck or VLT Interferometers. In fact, these instruments are 
expected to be able to measure image splittings as small as
$\sim 30\,\mu$as \citep{vlti}. Such measurements would allow a
direct measurement of the lens mass:
\begin{equation}
M = {\vp \that \theta_E c^2 \over 8G} \ .
\label{eq-Mtheta}
\end{equation}
The most challenging aspect of such measurements is the faintness
of source stars such as the MACHO-96-BLG-5 source, which is close to
the (rather uncertain) magnitude limit of the VLT Interferometer
\citep{vlti}.

Even if the images cannot be resolved, it may be possible
to measure the deflection of the image centroid
\citep{astrom1,astrom2,astrom3}
which is given by
\begin{equation}
\Delta\phi = \left( {u^2 +3\over \sqrt{u^2 + 4} } - u\right) \theta_E  \ .
\label{eq-cent}
\end{equation}
This can be measured by a very accurate astrometry
mission such as SIM \citep{boden,pac-astrom,gould-remnantm}. Once again,
however, the MACHO-96-BLG-5 source is a rather faint
target for SIM, but in this case, the measurement is
not so difficult because the amplitude of the centroid
motion is very much larger than SIM's sensitivity 
limit.

If it should turn out that some of the more massive lenses are located
very close to us, then it might be possible to directly observe 
the lensed images with HST. This is a realistic possibility for
the MACHO-99-BLG-22/OGLE-1999-BUL-32 event \citep{ogle-99b22} because its 
$\vpbold$ value is in the opposite quadrant from the events studied in this
paper. This gives a likelihood function with two peaks: one at a distance
of $\sim 500\,$pc for a lens in the disk and one at a distance of $\sim 6\,$kpc
for a bulge lens \citep{bennett-99b22}. The bulge lens solution predicts 
a mass of a few $\msun$, but the disk lens solution predicts a mass of
$> 100 \msun$ and a lensed image separation of $\sim 0.1$\rlap."

\subsection{Lens Detection and Source Proper Motion}
\label{sec-lensdetect}

Another method can be used to make a direct determination
of the lens mass for a bright lens star. If the
lens can be detected and the relative proper motion of the lens
with respect to the source is measured, then it is also possible
to determine the lens mass from the proper motion and microlensing
parallax parameters with the following formula:
\begin{equation}
M = {\vp \that^2 \mu c^2 \over 16G} \ ,
\label{eq-Mmu}
\end{equation}
where $\mu$ is the relative lens$-$source proper motion.
This technique has the advantage that the proper motion measurements
can be made many years after the peak magnification of the microlensing
event. The lens-source separation can reach the 50-100 mas range
within 5-10 years. Table~\ref{tbl-mmag} shows the predicted separations
and lens brightness contrasts for our six strong microlensing parallax events.
The columns are
(1) the MACHO event name, (2) the lens mass with $1\sigma$
errors, (3) a likely lens mass, $M_{rm MS}$, if the lens is on the
main sequence, (4) the lens distance, $D_{\ell -{\rm MS}}$ for a lens of
mass $M_{rm MS}$, (5) the predicted lens-source separation in June, 2003,
(6) the apparent V magnitude of the lenses, and
(7-11) the predicted contrast between the lens and source brightness in the
UBVI bands. Positive $\Delta$-mags.~imply that the source is brighter
than the lens, so lens detection is easiest for events that have
small or negative $\Delta$-mag.~values. With the exceptions of 
MACHO-96-BLG-5, which doesn't have a viable main sequence lens model,
all of the other lens
stars should be detectable if they are not stellar remnants.

When the lens can be detected, it should also be possible to constrain
the unlensed brightness of the source star, which will reduce the 
error bars on $\that$. Also, it should be possible to get very accurate
measures of the relative proper motion, $\mu$, as the lens moves
further from the source. Thus, the ultimate limits on the masses of the
lenses may come from the uncertainties in the $\vp$ values, which range
from $2-10\,$\%.

When the lenses are undetectable, it should still
be possible to measure the proper motion of the source star with
HST images separated by $\sim 5$ years. The proper motion can only
be measured with respect to the average of other, nearby stars because
extra-galactic reference sources are not easily identified in these
crowded Galactic bulge fields \citep{bul-pm,hst-bul-pm}.
Proper motion measurements of the microlensed source stars would
allow us to remove one degree of freedom from our likelihood analysis
and reduce the uncertainty in the implied lens distances and masses.
The proper motion distribution of the stars in the same field will
also allow us to test the Galactic models that are used for the
likelihood analysis, and so this should reduce the systematic
uncertainties in the lens distance and mass estimates.

\section{Discussion and Conclusions}
\label{sec-con}

We have performed microlensing parallax fits on the Galactic bulge
events detected by
the MACHO Collaboration with timescales of $\that \geq 140\,$days, and 
found six events with highly significant detections of the microlensing
parallax effect. Our analysis of the velocity distributions expected
for parallax microlensing events from different source star populations
suggests that source stars in the SGR Dwarf Galaxy might contribute to the
detectable microlensing parallax events, and inspection of the
source star color-magnitude diagrams indicates that two of our
microlensing parallax events have source stars which could be
SGR Dwarf red clump stars. However, radial velocity measurements
\citep{cook-spectra} indicate
that they are probably bulge sub-giant stars.

A likelihood analysis has been employed to estimate the distance and masses
of the lenses, and this indicates an average mass for our six lenses of
$2.7\,\msun$. Two of the lenses have masses large enough to imply that
they are probably massive stellar remnants: The mass estimates for the
MACHO-96-BLG-5 and MACHO-98-BLG-6 lenses are $M/\msun = 6{+10\atop -3}$ and 
$M/\msun = 6{+7\atop -3}$, respectively, which implies that both are 
likely to be black holes.
Together with MACHO-99-BLG-22/OGLE-1999-BUL-32 \citep{ogle-99b22}, 
these are the first black hole candidates that are truly black since
we have not seen any radiation from matter that is gravitationally bound
to the black hole.

Our likelihood analysis differs from that of \citet{agol} in that we 
compute the likelihood for the measured $\vpbold$ value whereas
\citet{agol} attempt to compute the likelihood of the measured values
of $\that$ as well as $\vpbold$. However, this requires that we input
the mass function of the lenses, and this has never been measured for
a complete sample of stellar remnants. 
Thus, the method of \citet{agol} can give misleading
results if the input mass function is not correct. Nevertheless, the
results of such an analysis are consistent with the results that we
have presented here. (Note that the preprint version of \citet{agol}
claimed an inconsistency with our results, but this was due to
an error in the computation of the likelihood function (Agol, private
communication).) For the MACHO-99-BLG-22/OGLE-1999-BUL-32 event, the
method of \citet{agol} does give potentially misleading results, however,
because the shape of the Likelihood function for this event
makes the results quite sensitive to the assumed black hole mass function
\citep{bennett-99b22}, which is, of course, unknown.

Similar events detected in the next few years may yield lens masses that are
measured much more precisely due to follow-up observations from ground-based
\citep{vlti} and space-based \citep{gould-remnantm} interferometers.
This will allow an unambiguous determination of the abundance 
and mass function of black hole and neutron star stellar remnants, although
it may be difficult to determine if $\sim 2\,\msun$ objects are black holes
or neutron stars. At present, there are three black hole microlens candidates
in the sample of 321 microlensing events that was the starting point for this
paper (although MACHO-99-BLG-22 is only identified as a strong black hole
candidate when OGLE data are included in the analysis 
\citep{ogle-99b22,bennett-99b22}).
This is about 1\% of the events, but far more than 1\% of the total
contribution to the microlensing optical depth. This suggests 
that the fraction of our Galaxy's stellar mass that is in the
form of black holes may be significantly larger than 1\%, which might
help to explain the observed excess of long timescale microlensing
events. However, we have not
made an accurate determination of our microlensing event detection efficiency 
for this data set, and the detection efficiency is certainly larger for
long timescale microlensing events than for short events. It is also possible
that one of these three lenses may not be a black hole, and so these
microlensing results may still be
consistent with models which predict that of order 1\% of the
Milky Way's stellar mass should be in the form of black holes
\citep{bb-bh,fryer01,gould-remnantm}. If all three of these events
are truly due to black hole lenses, then a black hole mass fraction as
high as $\sim 10$\% might be preferred. These results appear to indicate that
most stellar mass black holes do not reside in the X-ray binary systems
where they are most easily observed \citep{bh-mass}.

\acknowledgments

We thank Eric Agol for discussions regarding the use of a mass
prior in the likelihood analysis.
This work was supported, in part, by NASA through the Space Telescope
Science Institute (GO 8490) 
and through the NASA Origins Program (NAG5-4573). It was
also supported by the National Science Foundation grants program
(AST96-19575), and through the Office of Science
and Technology Centers (AST-8809616).
DM is supported by FONDAP Center for Astrophysic.
CWS thanks the Packard Foundation for the
generous support.  WJS is supported by a PPARC Advanced Fellowship.  CAN is
supported in part by a NPSC Graduate Fellowship.  TV and KG were supported in
part by the DOE.

\clearpage


\clearpage 

\clearpage 

\begin{figure}
\plottwo{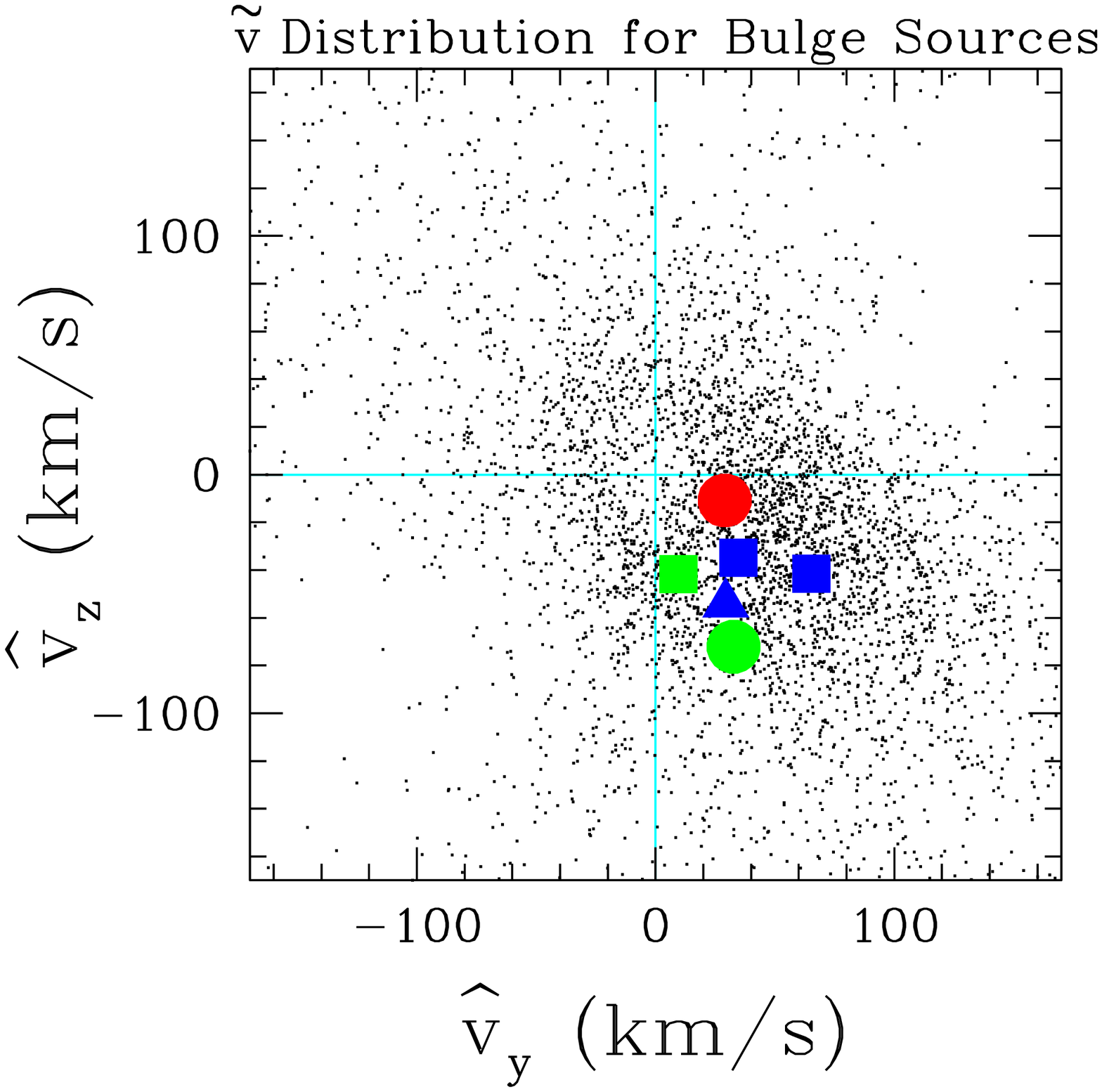}{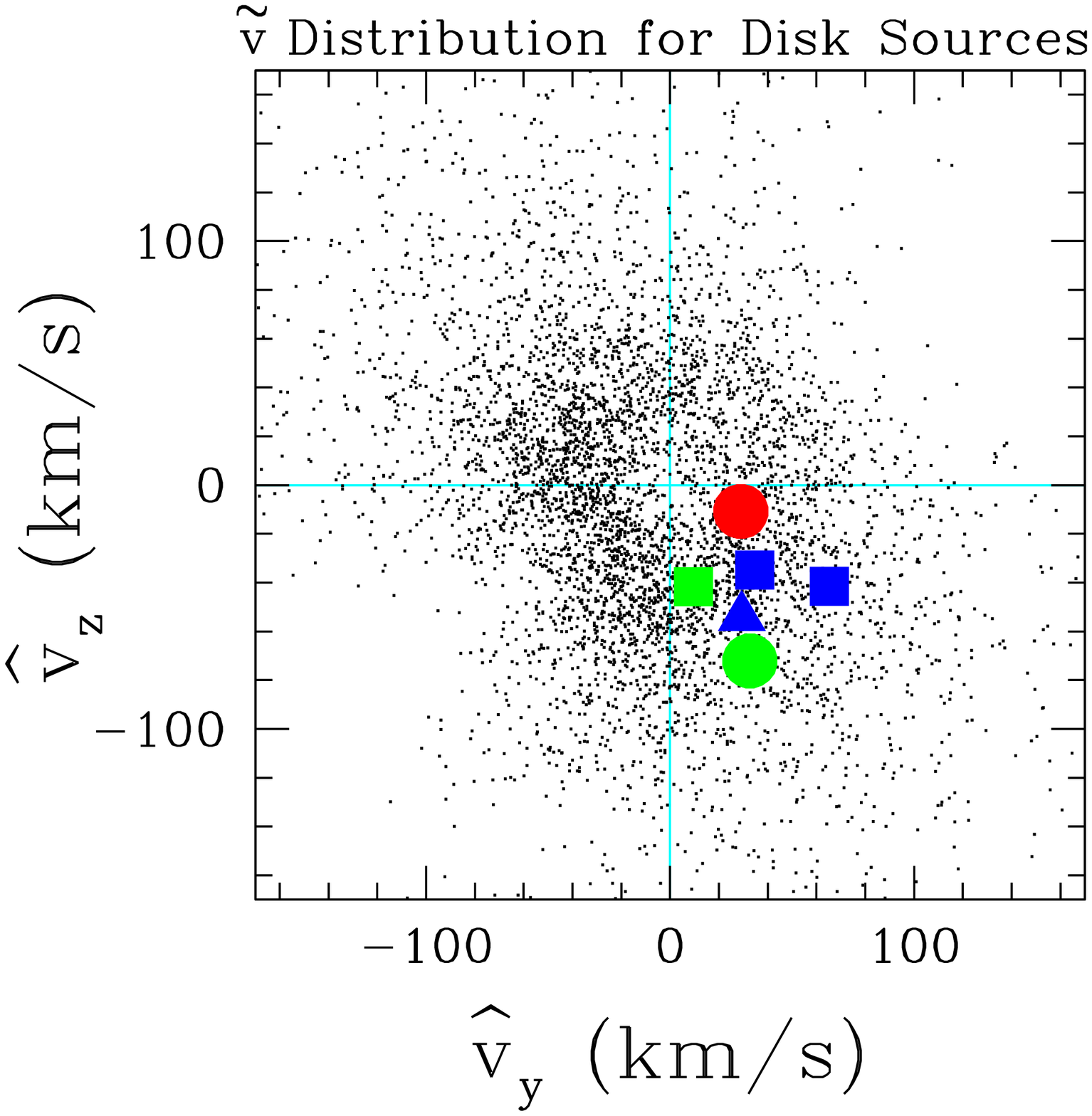}
\end{figure}

\clearpage 

\begin{figure}
\plottwo{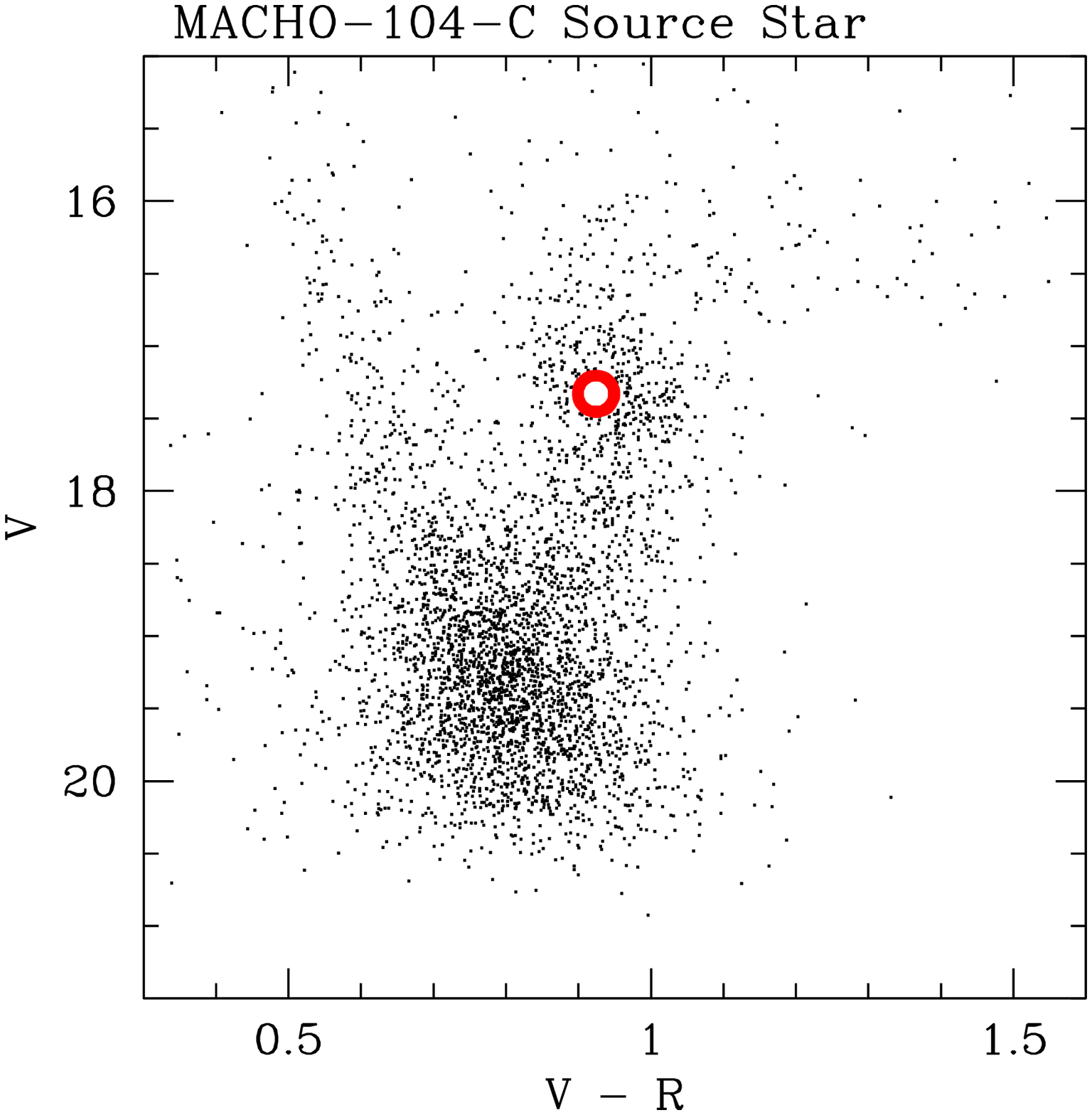}{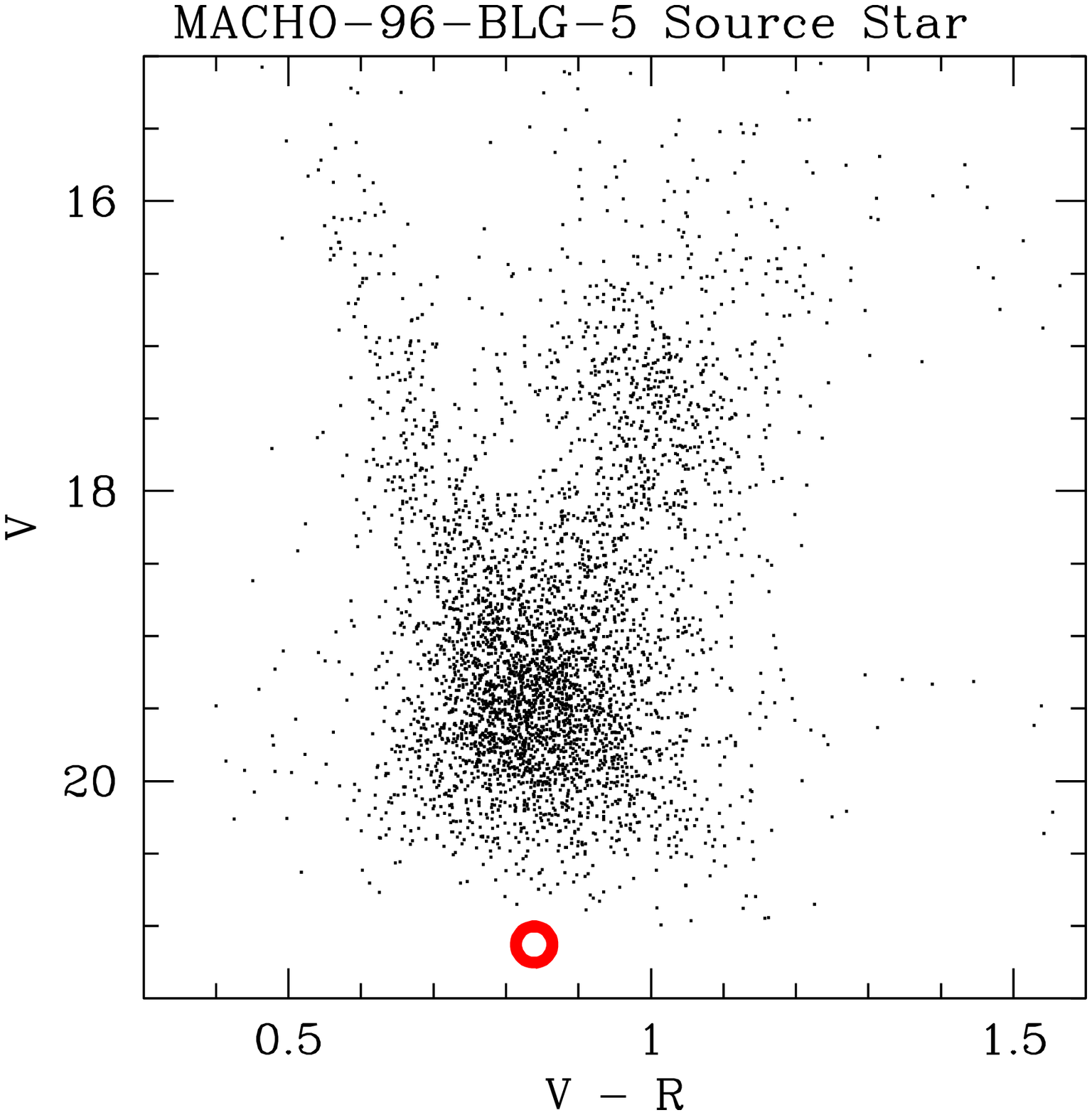}
\end{figure}
\begin{figure}
\plottwo{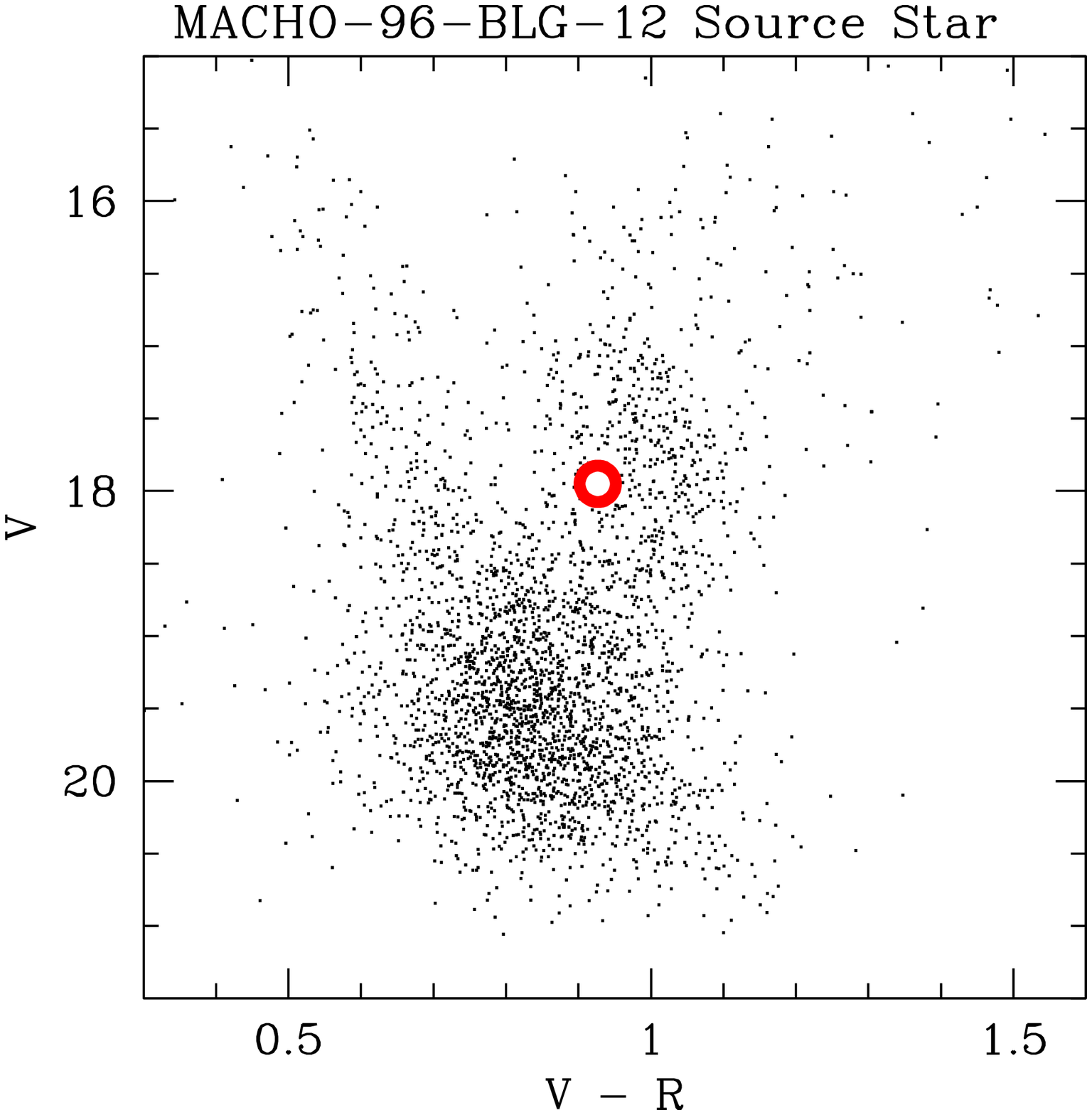}{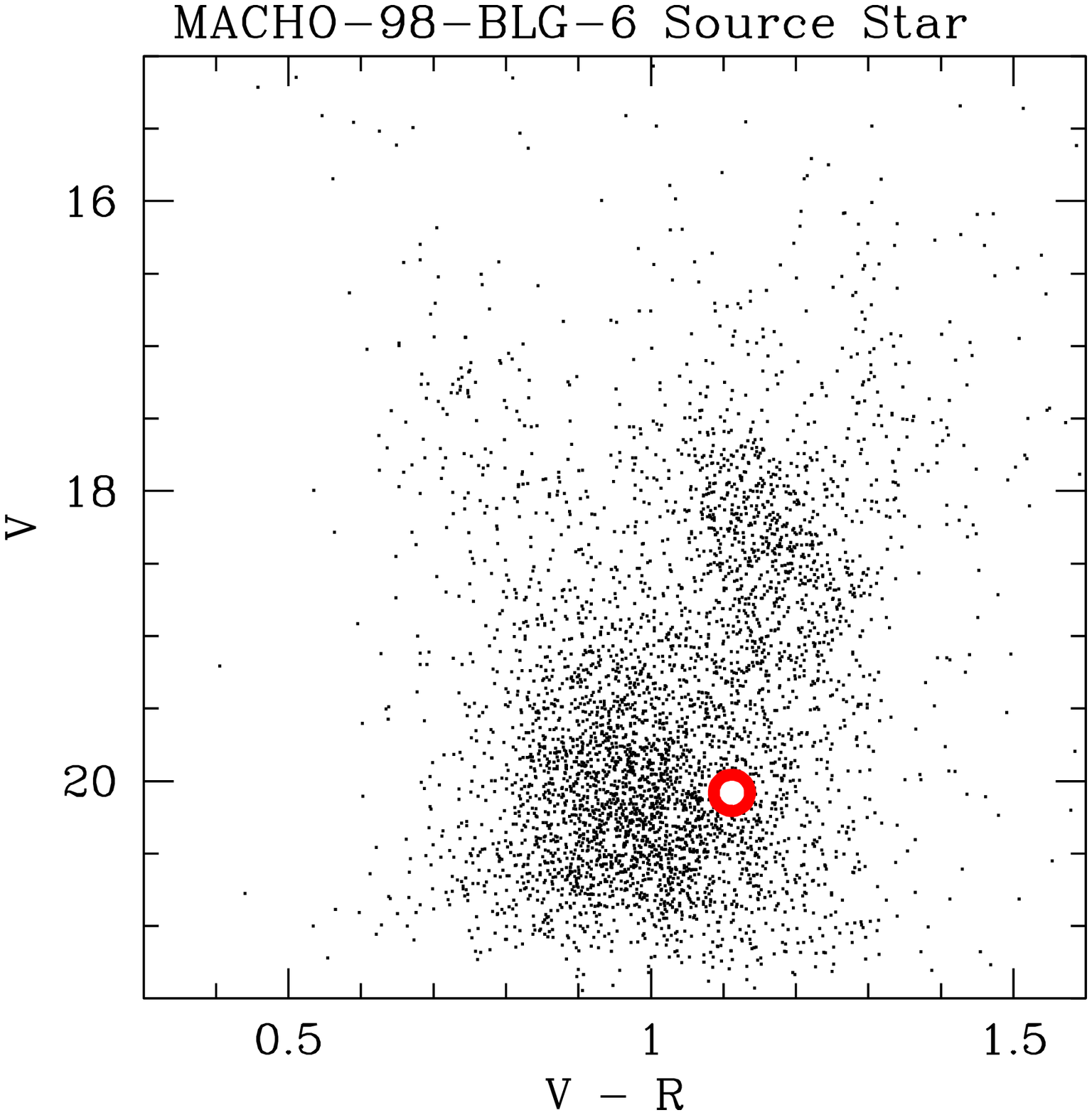}
\end{figure}

\clearpage 


\clearpage 

\clearpage






\clearpage

\begin{deluxetable}{llllcccc}
\tabletypesize{\small}
\tablecaption{Microlensing Parallax Event Coordinates\label{tbl-coords}}
\tablewidth{0pt}
\tablehead{
   &  &  &  & \multispan{2} Galactic & \multispan{2} Ecliptic \\
\colhead{Event Name} & \colhead{MACHO Star ID} & \colhead{RA(J2000)} &
\colhead{DEC(J2000)} & $l$ & $b$ & $\lambda$ & $\beta$ 
}
\startdata
MACHO-104-C    & 104.20251.50   & 18:03:34.0 & $-28$:00:19 & 2.797 & $-2.933$ & 270.790 & $-4.568$ \\
MACHO-96-BLG-5 & 104.20906.3973 & 18:05:02.5 & $-27$:42:17 & 3.219 & $-3.071$ & 271.119 & $-4.270$ \\
MACHO-96-BLG-12& 104.20382.803  & 18:03:53.2 & $-27$:57:36 & 2.871 & $-2.973$ & 270.861 & $-4.524$ \\
MACHO-98-BLG-6 & 402.48103.1719 & 17:57:32.8 & $-28$:42:45 & 1.526 & $-2.132$ & 268.762 & $-5.267$ \\
MACHO-99-BLG-1 & 121.22423.1032 & 18:08:50.0 & $-30$:31:56 & 1.138 & $-5.162$ & 271.917 & $-7.106$ \\
MACHO-99-BLG-8 & 403.47849.756  & 17:56:25.2 & $-29$:40:31 & 0.569 & $-2.401$ & 269.218 & $-6.237$ \\
 \enddata


\end{deluxetable}

\begin{deluxetable}{lcccc}
\tabletypesize{\small}
\tablecaption{Number of Observations\label{tbl-nobs}}
\tablewidth{0pt}
\tablehead{
\colhead{Event} & \colhead{MACHO-Red} & \colhead{MACHO-Blue} &
\colhead{CTIO} & \colhead{MPS}
}
\startdata
104-C    &  534 &  308 &    0 &    0 \\
96-BLG-5 &  558 & 1542 &  179 &    0 \\
96-BLG-12&  584 &  466 &  103 &    0 \\
98-BLG-6 &  952 & 1083 &   29 &  212 \\
99-BLG-1 &  343 &  260 &   11 &  153 \\
99-BLG-8 &  386 &  310 &  213 &  155 \\
 \enddata

\end{deluxetable}

\begin{deluxetable}{lcccc}
\tabletypesize{\scriptsize}
\tablecaption{Photometric Measurements\label{tbl-data}}
\tablewidth{0pt}
\tablehead{
\colhead{Event Name} & \colhead{Pass Band} & \colhead{time (MJD)} & 
\colhead{Magnitude} & \colhead{uncertainty}
}
\startdata
MACHO-104-C    & MACHO-Red  & 430.79500 & 14.0310  &  0.0221  \\
               &            & 438.78620 & 13.9090  &  0.0188  \\
               &            & 441.73940 & 13.8060  &  0.0163  \\
               &            & 442.74640 & 13.8120  &  0.0163  \\
               &            & 443.71500 & 13.7760  &  0.0172  \\
               &            & 446.72880 & 13.7400  &  0.0155  \\
               &            & 453.79520 & 13.5360  &  0.0182  \\
               &            & 459.71700 & 13.3230  &  0.0301  \\
               &            & 463.67350 & 13.2520  &  0.0172  \\
               &            & 463.67660 & 13.2570  &  0.0200  \\
               &            &   ...     & ...      &  ...     \\
               & MACHO-Blue & 430.79500 & 13.0240  &  0.0167  \\
               &            & 442.74640 & 12.7870  &  0.0157  \\
               &            & 443.71500 & 12.7820  &  0.0157  \\
               &            & 452.74020 & 12.5650  &  0.0157  \\
               &            & 453.79520 & 12.5320  &  0.0160  \\
               &            & 455.75470 & 12.4810  &  0.0157  \\
               &            & 457.78640 & 12.4300  &  0.0157  \\
               &            & 459.71700 & 12.3510  &  0.0200  \\
               &            & 463.67350 & 12.2290  &  0.0157  \\
               &            & 463.67660 & 12.2500  &  0.0167  \\
               &            &   ...     & ...      &  ...     \\
MACHO-96-BLG-5 & MACHO-Red  & 430.79500 & 16.2340  &  0.1299  \\
  & & 441.73940 & 16.3030  &  0.0726  \\
  & & 442.74640 & 16.4470  &  0.0814  \\
  & & 443.71500 & 16.2560  &  0.0942  \\
  & & 455.75470 & 16.2170  &  0.0952  \\
  & & 457.78640 & 16.3910  &  0.1110  \\
  & & 459.71700 & 15.8800  &  0.2295  \\
  & & 463.67350 & 16.3050  &  0.1538  \\
  & & 463.67660 & 15.9980  &  0.1796  \\
  & & 465.65960 & 16.4010  &  0.2574  \\
               &            &   ...     & ...      &  ...     \\
               & MACHO-Blue & 430.79500 & 17.0900  &  0.2155  \\
  & & 438.78620 & 17.2230  &  0.2066  \\
  & & 441.73940 & 17.1900  &  0.1031  \\
  & & 442.74640 & 17.3520  &  0.1140  \\
  & & 443.71500 & 17.2020  &  0.1388  \\
  & & 446.72880 & 17.4260  &  0.0580  \\
  & & 452.74020 & 17.1390  &  0.1577  \\
  & & 453.79520 & 17.0510  &  0.1747  \\
  & & 455.75470 & 17.2530  &  0.1547  \\
  & & 457.78640 & 17.3390  &  0.1647  \\
               &            &   ...     & ...      &  ...     \\
               & CTIO       & 1560.39200 & 15.5760  &  0.0550  \\
  & & 1560.39600 & 15.5580  &  0.0550  \\
  & & 1561.40400 & 15.5380  &  0.0493  \\
  & & 1561.40800 & 15.5300  &  0.0507  \\
  & & 1564.40100 & 15.5250  &  0.0465  \\
  & & 1564.40500 & 15.5670  &  0.0409  \\
  & & 1565.40300 & 15.4540  &  0.0437  \\
  & & 1565.40800 & 15.6850  &  0.0479  \\
  & & 1566.27800 & 15.5500  &  0.0409  \\
  & & 1566.28200 & 15.5280  &  0.0423  \\
               &            &   ...     & ...      &  ...     \\
MACHO-96-BLG-12& MACHO-Red  & 441.73940 & 14.2730  &  0.0319  \\
  & &  455.75470 & 14.2670  &  0.0382  \\
  & &  457.78640 & 14.2860  &  0.0382  \\
  & &  463.67660 & 14.3630  &  0.0668  \\
  & &  465.65960 & 14.2380  &  0.0542  \\
  & &  468.75260 & 14.3240  &  0.0736  \\
  & &  471.69700 & 14.2670  &  0.0259  \\
  & &  474.70090 & 14.3630  &  0.0336  \\
  & &  476.62860 & 14.2410  &  0.0336  \\
  & &  480.61710 & 14.2430  &  0.0301  \\
               &            &   ...     & ...      &  ...     \\
               & MACHO-Blue & 441.73940 & 15.2830  &  0.0382  \\
  & &  446.72880 & 15.5230  &  0.0200  \\
  & &  457.78640 & 15.2740  &  0.0513  \\
  & &  471.69700 & 15.3200  &  0.0336  \\
  & &  476.62860 & 15.3130  &  0.0447  \\
  & &  480.61710 & 15.2760  &  0.0409  \\
  & &  485.60160 & 15.3200  &  0.0513  \\
  & &  489.62250 & 15.2720  &  0.0590  \\
  & &  500.60370 & 15.2830  &  0.0428  \\
  & &  501.61230 & 15.2660  &  0.0372  \\
               &            &   ...     & ...      &  ...     \\
               & CTIO       & 1634.40000 & 13.3920  &  0.0194  \\
  & & 1634.40400 & 13.3740  &  0.0212  \\
  & & 1639.20200 & 13.3890  &  0.0243  \\
  & & 1640.14800 & 13.3640  &  0.0203  \\
  & & 1640.15200 & 13.3720  &  0.0222  \\
  & & 1653.21000 & 13.3330  &  0.0255  \\
  & & 1653.21400 & 13.3350  &  0.0255  \\
  & & 1661.15700 & 13.3090  &  0.0203  \\
  & & 1661.16100 & 13.3150  &  0.0194  \\
  & & 1668.01200 & 13.2860  &  0.0232  \\
               &            &   ...     & ...      &  ...     \\
MACHO-98-BLG-6 & MACHO-Red  & 1164.77190 & 15.8270  &  0.1180  \\
  & & 1168.75560 & 15.9450  &  0.1339  \\
  & & 1318.40870 & 15.8870  &  0.0590  \\
  & & 1319.40610 & 15.8430  &  0.0475  \\
  & & 1321.46010 & 16.1810  &  0.1448  \\
  & & 1323.46890 & 15.9430  &  0.1806  \\
  & & 1324.42410 & 16.2700  &  0.2814  \\
  & & 1325.41330 & 16.0320  &  0.1448  \\
  & & 1325.42910 & 15.9010  &  0.1220  \\
  & & 1326.46850 & 16.2210  &  0.1587  \\
               &            &   ...     & ...      &  ...     \\
               & MACHO-Blue & 1164.77190 & 17.1970  &  0.1856  \\
  & & 1168.75560 & 16.9490  &  0.1966  \\
  & & 1318.40870 & 17.1240  &  0.1170  \\
  & & 1319.40610 & 17.1600  &  0.0814  \\
  & & 1321.46010 & 17.1580  &  0.1210  \\
  & & 1323.46890 & 17.3430  &  0.2824  \\
  & & 1325.41330 & 17.1750  &  0.1796  \\
  & & 1325.42910 & 17.1360  &  0.1677  \\
  & & 1326.46850 & 17.2840  &  0.1468  \\
  & & 1327.43050 & 17.1130  &  0.0687  \\
               &            &   ...     & ...      &  ...     \\
               & CTIO       & 2303.38300 & 13.6610  &  0.0303  \\
  & & 2305.36700 & 13.6520  &  0.0243  \\
  & & 2308.40000 & 13.6610  &  0.0243  \\
  & & 2318.39500 & 13.5660  &  0.0290  \\
  & & 2323.41900 & 13.5270  &  0.0266  \\
  & & 2362.26600 & 13.2020  &  0.0437  \\
  & & 2438.03400 & 13.2050  &  0.0202  \\
  & & 2439.03100 & 13.2370  &  0.0278  \\
  & & 2440.03800 & 13.2320  &  0.0243  \\
  & & 2441.04900 & 13.2510  &  0.0266  \\
               &            &   ...     & ...      &  ...     \\
               & MPS        & 2306.71030 & 12.9630  &  0.0450  \\
  & & 2306.81750 & 13.0110  &  0.0395  \\
  & & 2308.55870 & 12.9080  &  0.0665  \\
  & & 2308.65670 & 13.1460  &  0.1479  \\
  & & 2308.71270 & 12.9020  &  0.0959  \\
  & & 2308.78110 & 12.9300  &  0.0395  \\
  & & 2316.68150 & 12.8670  &  0.0636  \\
  & & 2316.68370 & 12.9150  &  0.0535  \\
  & & 2328.59630 & 12.7660  &  0.0782  \\
  & & 2328.77350 & 12.7720  &  0.0593  \\
               &            &   ...     & ...      &  ...     \\
MACHO-99-BLG-1 & MACHO-Red  & 441.78840 & 15.3500  &  0.0301  \\
  & &  442.78270 & 15.3320  &  0.0251  \\
  & &  443.75950 & 15.3800  &  0.0428  \\
  & &  452.77860 & 15.3780  &  0.0504  \\
  & &  455.77750 & 15.3460  &  0.0400  \\
  & &  459.75610 & 15.3650  &  0.0419  \\
  & &  463.75080 & 15.3870  &  0.0629  \\
  & &  463.75430 & 15.3580  &  0.0600  \\
  & &  465.73690 & 15.6230  &  0.1637  \\
  & &  466.70240 & 15.3400  &  0.0687  \\
               &            &   ...     & ...      &  ...     \\
               & MACHO-Blue & 441.78840 & 16.2880  &  0.0428  \\
  & &  442.78270 & 16.2840  &  0.0327  \\
  & &  443.75950 & 16.2980  &  0.0629  \\
  & &  452.77860 & 16.3700  &  0.0795  \\
  & &  455.77750 & 16.3130  &  0.0648  \\
  & &  459.75610 & 16.2700  &  0.0854  \\
  & &  463.75080 & 16.1810  &  0.0972  \\
  & &  463.75430 & 16.1740  &  0.1041  \\
  & &  465.73690 & 16.7720  &  0.3813  \\
  & &  466.70240 & 16.4000  &  0.1349  \\
               &            &   ...     & ...      &  ...     \\
               & CTIO       & 2732.31900 & 13.3080  &  0.0232  \\
  & & 2733.20300 & 13.3090  &  0.0266  \\
  & & 2769.22400 & 13.4460  &  0.0194  \\
  & & 2778.22500 & 13.5420  &  0.0187  \\
  & & 2778.99300 & 13.5280  &  0.0222  \\
  & & 2784.13200 & 13.5770  &  0.0222  \\
  & & 2789.15200 & 13.6390  &  0.0522  \\
  & & 2794.96300 & 13.7050  &  0.0232  \\
  & & 2819.99200 & 13.9730  &  0.0255  \\
  & & 2820.99400 & 14.0080  &  0.0232  \\
               &            &   ...     & ...      &  ...     \\
               & MPS        & 2688.61020 & 12.6050  &  0.0328  \\
  & & 2688.71570 & 12.5720  &  0.1494  \\
  & & 2689.47020 & 12.6120  &  0.0841  \\
  & & 2689.55590 & 12.6540  &  0.0465  \\
  & & 2689.61810 & 12.6300  &  0.0507  \\
  & & 2689.71450 & 12.8030  &  0.1658  \\
  & & 2689.82380 & 12.6190  &  0.0382  \\
  & & 2690.56900 & 12.6450  &  0.0493  \\
  & & 2690.63600 & 12.6540  &  0.0221  \\
  & & 2690.66750 & 12.6410  &  0.0395  \\
               &            &   ...     & ...      &  ...     \\
MACHO-99-BLG-8 & MACHO-Red  & 1165.75610 & 11.7260  &  0.0194  \\
  & & 1166.76670 & 11.7530  &  0.0155  \\
  & & 1168.75920 & 11.7830  &  0.0155  \\
  & & 1318.41170 & 11.7830  &  0.0153  \\
  & & 1319.40950 & 11.7760  &  0.0153  \\
  & & 1323.47160 & 11.7490  &  0.0155  \\
  & & 1324.42880 & 11.7530  &  0.0157  \\
  & & 1325.41700 & 11.7610  &  0.0155  \\
  & & 1325.43250 & 11.7630  &  0.0155  \\
  & & 1326.47170 & 11.7480  &  0.0155  \\
               &            &   ...     & ...      &  ...     \\
               & MACHO-Blue & 1168.75920 & 13.4070  &  0.0172  \\
  & & 1323.47160 & 13.3740  &  0.0177  \\
  & & 1324.42880 & 13.3880  &  0.0188  \\
  & & 1325.43250 & 13.3850  &  0.0172  \\
  & & 1326.47170 & 13.4110  &  0.0167  \\
  & & 1327.43530 & 13.4010  &  0.0157  \\
  & & 1329.46640 & 13.3660  &  0.0163  \\
  & & 1330.42570 & 13.3340  &  0.0163  \\
  & & 1331.46440 & 13.4090  &  0.0447  \\
  & & 1342.45200 & 13.3370  &  0.0345  \\
               &            &   ...     & ...      &  ...     \\
               & CTIO       & 2630.39700 & 11.3820  &  0.0212  \\
  & & 2630.40000 & 11.3760  &  0.0194  \\
  & & 2630.40300 & 11.3710  &  0.0255  \\
  & & 2630.40600 & 11.3960  &  0.0243  \\
  & & 2632.39400 & 11.3020  &  0.0290  \\
  & & 2632.39600 & 11.3220  &  0.0355  \\
  & & 2632.40000 & 11.3240  &  0.0355  \\
  & & 2632.40200 & 11.3840  &  0.0342  \\
  & & 2639.37200 & 11.3020  &  0.0243  \\
  & & 2639.37500 & 11.2510  &  0.0203  \\
               &            &   ...     & ...      &  ...     \\
               & MPS        & 2688.60420 & 12.4140  &  0.0179  \\
  & & 2688.60530 & 12.4340  &  0.0186  \\
  & & 2688.60680 & 12.4310  &  0.0186  \\
  & & 2688.71320 & 12.3960  &  0.0493  \\
  & & 2689.46830 & 12.4360  &  0.0232  \\
  & & 2689.55290 & 12.3980  &  0.0221  \\
  & & 2689.61620 & 12.4230  &  0.0232  \\
  & & 2689.71320 & 12.4310  &  0.0202  \\
  & & 2690.47590 & 12.3160  &  0.0564  \\
  & & 2690.56730 & 12.4250  &  0.0202  \\
               &            &   ...     & ...      &  ...     \\
 \enddata
\tablecomments{Complete data set available in electronic version.
               ${\rm MJD} = {\rm JD} - 248623.5\,$ days.}
\end{deluxetable}

\begin{deluxetable}{lcccc}
\tabletypesize{\small}
\tablecaption{Photometric Calibration Coefficients\label{tbl-calib}}
\tablewidth{0pt}
\tablehead{
\colhead{Event Name} & \colhead{$a$} & \colhead{$b$} & 
\colhead{$c$} & \colhead{$d$}
}
\startdata
MACHO-104-C    & $0.8176$ & $0.1824$ & $0.1828$ & $0.8172$ \\
MACHO-96-BLG-5 & $0.8076$ & $0.1924$ & $0.1804$ & $0.8196$ \\
MACHO-96-BLG-12& $0.8176$ & $0.1824$ & $0.1828$ & $0.8172$ \\
MACHO-98-BLG-6 & $0.8191$ & $0.1809$ & $0.1829$ & $0.8171$ \\
MACHO-99-BLG-1 & $0.8169$ & $0.1831$ & $0.1826$ & $0.8174$ \\
MACHO-99-BLG-8 & $0.8188$ & $0.1812$ & $0.1829$ & $0.8171$ \\
 \enddata

\end{deluxetable}

\begin{deluxetable}{lllllcccclrr}
\tabletypesize{\scriptsize}
\tablecaption{Microlensing Parallax Fit Parameters \label{tbl-fitpar}}
\tablewidth{0pt}
\tablehead{
\colhead{Event} & \colhead{$f_{MR}$} & \colhead{$f_{MB}$} &
\colhead{$f_{CTIO}$} & \colhead{$f_{MPS}$} &  
\colhead{$t_0$ (MJD)} & \colhead{$\umin$} & 
\colhead{$\that$ (days)}   &  \colhead{$\vp$ (km/sec)} &
\colhead{$\theta$} & \colhead{${\chi^2\over {\rm(dof)}}$} & 
\colhead{$\Delta\chi^2$}
}
\startdata
104-C    & $1.00(1)$ & $0.99(2)$ & & & $508.3(6)$ & $0.15(1)$ & $220(2)$ & $77(4)$ & $-1.08(7)$ & $1.47$ & $1051$ \\
96-BLG-5 & $0.12(3)$ & $0.12(3)$ & $0.13(3)$ & & $1763(1)$ & $0.018(6)$ & $2000(500)$ & $30.9(1.3)$ & $-0.84(6)$ & $1.58$ & $2395$ \\
\ \ (HST)& $0.28(1)$ & $0.30(1)$ & ${\bf 0.33}$ & & $1767(1)$ & $0.048(6)$ & $970(20)$ & $30.9(1.3)$ & $-0.87(7)$ & $1.59$ & $2371$ \\
               & $0.31(1)$ & $0.33(1)$ & $0.37$ & & $1768(1)$ & $0.054(7)$ & $900(20)$ & $31.0(1.3)$ & $-0.88(8)$ & $1.59$ & $2363$ \\
96-BLG-12& $0.87(2)$ & $0.89(3)$ & $0.90(2)$ & & $1743.4(3)$ & $-0.11(2)$ & $294(5)$ & $47.5(1.3)$ & $-1.23(9)$ & $2.11$ & $5914$ \\
98-BLG-6 & $0.65(14)$ & $0.60(13)$ & $0.68(15)$ & $0.66(13)$ & $2388(3)$ & $0.16(4)$ & $490(50)$ & $79(5)$ & $-1.7(2)$ & $1.20$ & $802$ \\
99-BLG-1 & $0.96(9)$ & $0.98(10)$ & $1.0(1)$ & $0.97(7)$ & $2712(1)$ & $0.23(4)$ & $231(13)$ & $43.9(9)$ & $-1.85(2)$ & $1.54$ & $1706$ \\
99-BLG-8 & $0.75(12)$ & $0.73(12)$ & $0.76(12)$ & $0.79(13)$ & $2732.1(4)$ & $0.17(1)$ & $240(20)$ & $62(5)$ & $-1.53(3)$ & $2.34$ & $2280$ \\
 \enddata


\tablecomments{${\rm MJD} = {\rm JD} - 248623.5\,$ days.}
\end{deluxetable}

\begin{deluxetable}{lllllccccc}
\tabletypesize{\scriptsize}
\tablecaption{Microlensing Parallax Likelihood Mass Estimates \label{tbl-mass}}
\tablewidth{0pt}
\tablehead{
 & & & & \multispan{5}\hfil\  \ \
     {Confidence Levels $P(M/\msun < N)$ }\hfil  \\
\colhead{Event} & \colhead{location} & \colhead{$f_{MR}$}   & 
\colhead{$\that$ (days)}   &  \colhead{$\vhat$ (km/sec)} &
\colhead{$P=5\%$}  & \colhead{$P=16\%$} & \colhead{$P=50\%$} &
\colhead{$P=84\%$}     & \colhead{$P=95\%$}
}
\startdata
104-C    & bulge & $1.00(1)$ & $220(2)$ & $77(4)$ & {\bf 0.35} & {\bf 0.62} & {\bf 1.15} & {\bf 2.2}  & {\bf 3.94}\\
96-BLG-5 & bulge & $0.12(3)$&$2000(500)$ & $30.9(1.3)$& $7.2$ & $12.8$ & $27$   & $69$   & $160$\\
\ \ (HST)       & bulge & $0.28(1)$ & $970(20)$ & $30.9(1.3)$& {\bf 1.64} & {\bf 2.93} & {\bf 6.3}  & {\bf 15.8} & {\bf 37} \\
               & bulge & $0.31(1)$ & $900(20)$ & $31.0(1.3)$& $1.41$& $2.53$ & $5.4$  & $13.6$ & $31$ \\
96-BLG-12& bulge & $0.87(2)$ & $294(5)$ & $47.5(1.4)$ & {\bf 0.33} & {\bf 0.62} & {\bf 1.29} & {\bf 3.1}  & {\bf 6.7} \\
98-BLG-6 & bulge & $0.65(14)$ & $490(50)$ & $79(6)$ & {\bf 0.94} & {\bf 2.6}  & {\bf 5.7} & {\bf 12.5} & {\bf 24} \\
               & SGR   & $0.65(14)$ & $490(50)$ & $79(6)$ & 1.23 & 1.61 & 2.52 & 4.2  & 6.7 \\
99-BLG-1 & bulge & $0.96(9)$ & $219(9)$ & $42.9(9)$ & {\bf 0.14} & {\bf 0.29} & {\bf 0.68} & {\bf 1.86} & {\bf 4.6} \\
               & SGR   & $0.96(9)$ & $219(9)$ & $42.9(9)$ & 0.10 & 0.16 & 0.31 & 0.70 & 1.51 \\
99-BLG-8 & bulge & $0.75(12)$ & $240(20)$ & $62(5)$ & {\bf 0.27} & {\bf 0.56} & {\bf 1.19} & {\bf 2.78}& {\bf 6.0} \\
 \enddata


\tablecomments{Bold-faced type indicates the parameters that are considered to 
be most likely.  Event 96-BLG-5 has parameters for three different fits listed.
The first fit is the fit with no constraint on the source brightness, while the
second and third fits have the lensed flux fixed to a value based upon our
HST observations. The fit labeled HST is the best fit, while the third fit is
provided to indicate the effect of the source flux uncertainty on the mass
limits. For events 98-BLG-6 and 99-BLG-1, mass estimates based upon bulge
and SGR sources are presented. In all cases, it is most likely that the
source star is in the bulge.}

\end{deluxetable}

\begin{deluxetable}{llclrrrrrr}
\tabletypesize{\scriptsize}
\tablecaption{Mass \& Magnitude Estimates for the MACHO Microlensing Parallax Events\label{tbl-mmag}}
\tablewidth{0pt}
\tablehead{
\colhead{Event} & \colhead{$M/\msun$} &
\colhead{$M_{\rm MS}/\msun$} & \colhead{$D_{\ell-{\rm MS}}$} &
\colhead{sep-MS} & \colhead{$V_s$} & \colhead{$\Delta I_{\ell s}$} &
\colhead{$\Delta V_{\ell s}$} & \colhead{$\Delta B_{\ell s}$} &
\colhead{$\Delta U_{\ell s}$} 
}
\startdata
104-C    & $1.1{+1.1\atop -0.5}$ & $0.74$ & $2.7\kpc$ & $40\,$mas &
           $17.3$ & $3.5$ & $3.5$ & $3.5$ & $3.2$ \\ 
96-BLG-5 & $6{+10\atop -3}$ & - & - & - & - & - & - & - & - \\
96-BLG-12& $1.3{+1.8\atop -0.7}$ & $0.75$ & $2.0\kpc$ & $28\,$mas &
           $18.0$ & $2.1$ & $2.2$ & $2.2$ & $2.3$ \\ 
98-BLG-6 & $2.5{+1.7\atop -0.9}$ & $0.88$ & $5.7\kpc$ & $5\,$mas &
           $20.1$ & $2.2$ & $1.9$ & $1.6$ & $1.1$ \\ 
99-BLG-1 & $0.7{+1.2\atop -0.4}$ & $0.40$ & $1.7\kpc$ & $17\,$mas &
           $18.9$ & $1.8$ & $3.2$ & $3.6$ & $3.9$ \\ 
99-BLG-8 & $1.2{+1.6\atop -0.6}$ & $1.2$  & $1.6\kpc$ & $25\,$mas &
           $16.3$ & $1.3$ & $0.7$ & $-0.3$ & $-1.1$ \\ 
 \enddata


\tablecomments{These are the parameters of the ``most likely" main
sequence star lenses for our best microlensing parallax events. For
MACHO-96-BLG-5, a main sequence lens is ruled out. }

\end{deluxetable}

\end{document}